\documentclass[twocolumn]{aastex6}
\usepackage{graphicx}
\usepackage{amsmath}
\usepackage{natbib}

\usepackage{color}

\begin{document} 
\title{A test of the high-eccentricity migration scenario for close-in planets}
\author{Steven Giacalone\altaffilmark{1}, Titos Matsakos\altaffilmark{2}, and Arieh K\"onigl\altaffilmark{2}}
\altaffiltext{1}{Department of Physics, The University of Chicago, Chicago, IL 60637, USA}
\altaffiltext{2}{Department of Astronomy \& Astrophysics and The Enrico Fermi Institute,
The University of Chicago, Chicago, IL 60637, USA}
\shortauthors{Giacalone, Matsakos, \& K\"onigl}
\shorttitle{A test of the high-eccentricity migration scenario}
\begin{abstract}
In the high-eccentricity migration (HEM) scenario, close-in planets reach the vicinity of the central star on high-eccentricity orbits that become circularized---with a concomitant decrease in the semimajor axis---through a tidal interaction with the star. Giant planets that arrive with periastron distances that are smaller than the Roche limit $a_\mathrm{R}$ lose their gaseous envelopes, resulting in an inner edge to the surviving planets' distribution. The observational evidence for this effect, while extensive, is nonetheless somewhat ambiguous because of the effect of tidal orbital decay. Here we consider another key prediction of the HEM scenario---the existence of a spatial eccentricity gradient near the location where the circularization time becomes comparable to the planet's age for typical parameters. Previous studies already found evidence for this gradient and demonstrated that its properties are consistent with the circularization process being dominated by tidal dissipation in the planet (encapsulated by the tidal quality factor $Q^\prime_\mathrm{p}$). Our work extends these treatments by constructing explicit model distributions for comparison with the data and by carrying out backward-in-time integrations using observed system parameters. We show that circularization generally occurs outside the distribution's inner edge (which defines the boundary of the so-called sub-Jovian desert) and that typically $Q^\prime_\mathrm{p}\approx10^6$ in the circularization zone (to within a factor of 3). We also find tentative evidence for an eccentricity gradient in lower-mass planets, indicating that formation through HEM may be relevant down to Neptune scales.
\end{abstract}
\keywords{
  planets and satellites: dynamical evolution and stability --- 
  planets and satellites: formation --- planet-star interactions
}
\maketitle
\section{Introduction}
\label{sec:intro}
The growing number of observed close-in exoplanets (planets with orbital periods $P_\mathrm{orb}\lesssim 10$\,days) has motivated researchers to look for trends in the distribution of their physical and orbital parameters that might help clarify the origin of these planets and the nature of their interaction with the host star. In an early study of this type, \citet[][hereafter PHMF11]{Pont+11} drew attention to two such trends for giant planets: the prevalence of circular orbits for very short periods, and the inverse correlation between the planet's mass $M_\mathrm{p}$ and $P_\mathrm{orb}$ for the closest planets. Under the prevailing view, wherein giant planets form beyond the water--ice line, the first of these trends has two distinct potential interpretations: in the disk migration scenario, the observed planets reach the central star by drifting inward through the protoplanetary disk on nearly circular orbits \citep[e.g.,][]{Lin+96}, whereas in the high-eccentricity migration (HEM) picture they arrive on high-eccentricity orbits that become tidally circularized when the planets approach the star \citep[e.g.,][]{RasioFord96}. In this connection, PHMF11 identified an unambiguous transition from eccentric to circular orbits on going from long to short orbital periods (for a given value of $M_\mathrm{p}$) and from high-mass to low-mass  planets (for a given value of $P_\mathrm{orb}$), and pointed out that this behavior is consistent with the expected outcome of a circularization process that is dominated by tidal dissipation in the planet \citep[see also][]{Husnoo+12}. These results were confirmed in a recent study by \citet{Bonomo+17}.

In trying to interpret the second identified trend---the pile-up of the shortest-period planets in such a way that those with higher values of $M_\mathrm{p}$ have lower values of $P_\mathrm{orb}$---PHMF11 speculated that tidal circularization and the stopping mechanism of close-in planets might be related. However, this mass--period relation was subsequently recognized to be part of a more general feature in the $P_\mathrm{orb}$--$M_\mathrm{p}$ plane: a nearly empty area, outlined roughly by two oppositely sloped lines, in the region of sub-Jupiter-mass planets on short-period orbits \citep[e.g.][]{SzaboKiss11,BeaugeNesvorny13,Mazeh+16}. \citet[][hereafter MK16]{MatsakosKonigl16} showed that this feature (dubbed the sub-Jovian desert) can be interpreted in terms of HEM, with the two distinct segments of the desert's boundary reflecting the different slopes of the empirical mass--radius relation for small and large planets \citep[e.g.,][]{Weiss+13}.  A plausible physical origin for the boundary is the Roche limit $a_\mathrm{R}$, the distance from the star where the planet starts to be tidally disrupted. If the planet arrives on an orbit with initial (subscript 0) semimajor axis $a_0$ and eccentricity $e_0$, its distance of closest approach will be $a_\mathrm{per,0}=(1-e_0)\,a_0$ and it will circularize (assuming conservation of orbital angular momentum) at $a_\mathrm{cir}=(1+e_0)\,a_\mathrm{per,0}$ (i.e., at $\simeq 2\,a_\mathrm{per,0}$ for a highly eccentric orbit; \citealt{RasioFord96}). The upper boundary of the sub-Jovian desert was already interpreted in this way by \citet{FordRasio06}, although the data available at the time was insufficient for a definitive model fit.\footnote{As was pointed out by MK16, the observed shape of the desert's upper boundary is not adequately reproduced unless one also takes into account the tidal dissipation in the star, which, on a timescale much longer than the circularization time, causes the planet's orbit to decay. In a previous study, \citet{ValsecchiRasio14} interpreted the finding of giant planets with semimajor axes $< 2\,a_\mathrm{R}$ in terms of orbital decay of this type.} The high-$e_0$ orbit could originate in a sudden planet--planet scattering event or in a slower interaction such as Kozai migration (involving either a stellar or a planetary companion) or secular chaos. In the latter case \citet{WuLithwick11} suggested that the planet's inward drift might be arrested at the location where the rate at which its longitude of pericenter precesses due to a secular interaction with a more distant planet is equal to the orbit-averaged precession rate associated with the tidal quadrupole induced on the planet by the star. The locus of circularization radii in the $P_\mathrm{orb}$--$M_\mathrm{p}$ plane is, however, similar in this case to the one obtained by setting $a_\mathrm{per,0}=a_\mathrm{R}$. Note that the planet's stopping mechanism is \emph{not} directly related to the circularization process in either of these two explanations of the pile-up. However, in order for the data to be compatible with the prediction $a_\mathrm{cir} \approx 2 \,a_\mathrm{per,0}$, the circularization radius $r_\mathrm{cir}$ (obtained by equating the planet's circularization time to the time that has elapsed since its arrival at the stellar vicinity) must exceed $a_\mathrm{cir}$. As we demonstrate in Section~\ref{sec:results}, this condition is typically satisfied for planets in the pile-up zone.

The observational support for the role of the HEM mechanism in shaping the spatial distribution of close-in planets has so far been based primarily on the apparent paucity of planets with semimajor axes $a\lesssim 2\,a_\mathrm{R}$ \citep[e.g.,][]{RasioFord96,Matsumura+10,ValsecchiRasio14} and on the corresponding dearth (the sub-Jovian desert) in the period--mass plane (\citealt{FordRasio06}; MK16). While the observational evidence for this effect is strong, it is not entirely unambiguous on account of the (already noted) additional orbital evolution induced by tidal dissipation in the star. In this paper we consider a complementary observational test of this scenario, the expected gradient in planet eccentricities in the vicinity of the locus of the circularization radii in the period--mass plane. The existence of such a gradient was already demonstrated in PHMF11, but here, in addition to updating the database, we compare it explicitly with the predictions of the HEM model. We describe our modeling approach in Section~\ref{sec:model}, present results in Section~\ref{sec:results}, discuss the main implications in Section~\ref{sec:discuss}, and summarize in Section~\ref{sec:conclude}. In a separate paper \citep[][hereafter KGM17]{Konigl+17} we use the model employed in this work to study the fate of high-mass planets that arrive by HEM and end up crossing the Roche limit, which results in the loss of their gaseous envelopes: we argue that the remnant rocky cores of these planets can plausibly account for the recently identified population of dynamically isolated hot Earths \citep{SteffenCoughlin16}.
\section{Modeling approach}
\label{sec:model}
Our treatment is based on the formulation presented in MK16, whose work was similarly concerned with the properties of close-in planets that arrive by HEM and undergo orbital circularization through internal tidal dissipation. That paper examined the shape of the boundary of the sub-Jovian desert in the $P_\mathrm{orb}$--$M_\mathrm{p}$ plane under the assumption that planets reach the vicinity of the Roche limit $a_\mathrm{R}$ (or, alternatively, the point of closest approach in the secular-chaos model of \citealt{WuLithwick11}) with $e_0 \approx 1$ and that their orbits then undergo an effectively instantaneous circularization. In contrast with that work, in which only the post-circularization orbital evolution of planets due to tidal dissipation in the star was calculated, in this paper we also account explicitly for tidal dissipation in the planets and we consider its effect on orbital circularization for a range of initial eccentricities and without assuming a priori that this process always runs to completion. We carry out Monte Carlo simulations using the same distributions of $P_\mathrm{orb,0}$, $R_\mathrm{p}$, $M_\mathrm{p}$, and $t_\mathrm{arr}$ (the planet arrival time at the stellar vicinity) as in MK16, but we update their choices for the distributions of radii and masses of large planets as well as of  $t_\mathrm{age}$ (the system's age) using data downloaded from the Extrasolar Planets Encyclopedia database at \textit{exoplanet.eu}.

Any given system is specified by six parameters: $P_\mathrm{orb,0}$, $e_0$, $R_\mathrm{p}$, $M_\mathrm{p}$, $t_\mathrm{age}$, and $t_\mathrm{arr}$. We assume that the planets that reach the stellar vicinity by HEM originate in the $P_\mathrm{orb,0}$ range [10,100]\,days with $e_0$ in the range [0.5,0.9]. Though consistent with our current understanding of how the HEM process operates in real systems \citep[e.g.,][]{Dawson+15,PetrovichTremaine16}, these ranges are not meant to represent any particular physical model and are chosen for illustrative purposes only. For $P_\mathrm{orb,0}$ we adopt the empirical distribution $\partial{f}/\partial\log{P_\mathrm{orb,0}} \propto P_\mathrm{orb,0}^{0.47}$ given in \citet[][]{Youdin11}, whereas for the $e_0$ distribution we adopt the form $\partial{f}/\partial e_0 = \textit{constant}$ (which corresponds to the steady-state distribution obtained by \citealt{PetrovichTremaine16} and \citealt{Antonini+16} for planets that undergo eccentricity oscillations due to secular gravitational interactions with an outer companion, as well as to the ``coplanar'' distribution obtained in the secular-chaos model and shown in figure~4 of \citealt{WuLithwick11}).\footnote{The results presented in this paper are not sensitive to the details of the $P_\mathrm{orb,0}$ and $e_0$ distributions. However, in KGM17 we examine the dependence of the fraction of planets that end up crossing the Roche limit on the form of the $e_0$ distribution and on the maximum value of $e_0$.} The values of $R_\mathrm{p}$ are also sampled from an empirical distribution given in \citet{Youdin11}, $\partial{f}/\partial\log{R_\mathrm{p}} \propto R_\mathrm{p}^{-0.66}$. As in MK16, we distinguish between small and large planets, separated at $R_\mathrm{p} = 12\, R_\earth$. For the small planets we adopt 
\begin{equation}
M_\mathrm{p} = \left(\frac{R_\mathrm{p}}{R_\earth}\right)^2 M_\earth\, ,\quad\quad R_\mathrm{p}<12\,R_\earth
\label{eq:RpMp}
\end{equation}
\citep[see][]{Weiss+13}. In the case of the large planets---for which $R_\mathrm{p}$ is nearly independent of $M_\mathrm{p}$---we resample $R_\mathrm{p}$ from the interval [9,20]\,$R_\earth$ and independently sample $M_\mathrm{p}$ from the interval [0.3,10]\,$M_\mathrm{J}$ using the \textit{exoplanet.eu} database (see Appendix~\ref{app:distributions}; note that this was done in MK16 using the data originally compiled by \citealt{Weiss+13}).\footnote{The radii of giant planets evidently depend on the incident flux from the host star \citep[e.g.,][]{Laughlin+11} and could thus vary systematically with orbital period. We do not explicitly account for this dependence since it was found to be fairly weak \citep[e.g.,][]{Weiss+13} and because the orbital circularization zone that we investigate corresponds to a rather narrow range of $P_\mathrm{orb}$ values.} In another modification of the MK16 implementation, we replace the age distribution given in \citet{WalkowiczBasri13}---which was determined by applying gyrochronology relationships to comparatively rapidly rotating stars and is therefore biased toward young systems---by an empirical distribution obtained from \textit{exoplanet.eu}. We restrict attention to the $P_\mathrm{orb}$ interval [2.5,7]\,days (where the upper bound defines the regime of hot Jupiters and the lower limit roughly corresponds to the distance from the star below which the effect of tidal orbital decay becomes significant) and consider separately the age distributions for small and large planets (see Appendix~\ref{app:distributions}). Finally, we assume a uniform distribution in $\log{t_\mathrm{arr}}$ for the arrival times, with $t_\mathrm{arr}$$\in$[0.01,10]\,Gyr (see MK16).

After sampling for $P_\mathrm{orb,0}$ and $e_0$, one can determine $a_\mathrm{per,0}(a_0,e_0)$ and $a_\mathrm{cir}(a_0,e_0)$ (setting $a_0 =(GM_*P_\mathrm{orb,0}^2/4\pi^2)^{1/3}$, where $G$ is the gravitational constant and $M_*$ is the stellar mass). These values can, in turn, be used to select the systems that are relevant to the present calculation. In this work we only consider the Roche-limit interpretation of the sub-Jovian desert boundary, taking the lower bound on a planet's initial periastron distance to be given by 
\begin{eqnarray}
\label{eq:RL}
a_\mathrm{R}&=&q(M_*/M_\mathrm{p})^{1/3}\,R_\mathrm{p}\\
&=& 0.016 \left(\frac{q}{3.46}\right) \left(\frac{M_*} {M_{\mathrm{\sun}}}\right)^{1/3} \left(\frac{M_{\mathrm{p}}}{M_{\mathrm{J}}}\right)^{-1/3} \left(\frac{R_{\mathrm{p}}}{R_{\mathrm{J}}}\right) \mathrm{au}\nonumber
\end{eqnarray}
(where the normalization of the coefficient $q$ is based on the results of MK16). The systems whose evolution we follow are defined by the requirements $a_\mathrm{per,0}>a_\mathrm{R}$ and $P_\mathrm{orb}(a_\mathrm{cir})\lesssim 7$\,days. We simplify the evolution equations by assuming that the orbital angular momentum vector is aligned with the spin vector of the star as well as with that of the planet, and that the planet's rotation period does not change with time (corresponding to pseudosynchronicity). We also neglect the time variation of the stellar rotation period $P_*$ (which we assume to be distributed uniformly in the interval [5,10]\,days). The evolution equations are then given by
\begin{equation}
\label{eq:a_t}
\begin{split}
  \frac{da}{dt}=\frac{9}{Q^\prime_\mathrm{p}}\left (\frac{GM_*}{a^3}\right )^{1/2}\frac{M_*}{M_{\mathrm{p}}}\frac{R_{\mathrm{p}}^5}{a^4}(1-e^2)^{-15/2}\\ \times\left[\frac{(\textit{f}_2(e^2))^2}{\textit{f}_5(e^2)}-\textit{f}_1(e^2) \right]\\
+\,\frac{9}{Q^\prime_*}\left (\frac{GM_*}{a^3}\right )^{1/2}\frac{M_\mathrm{p}}{M_*}\frac{R_*^5}{a^4}(1-e^2)^{-15/2}\\ \times\left[\textit{f}_2(e^2)\frac{P_\mathrm{orb}}{P_*}(1-e^2)^{3/2}-\textit{f}_1(e^2) \right]
\end{split}
\end{equation}
and
\begin{equation}
\label{eq:e_t}
\begin{split}
  \frac{de}{dt}=\frac{81}{2\,Q^\prime_\mathrm{p}}\left (\frac{GM_*}{a^3}\right )^{1/2}\frac{M_*}{M_{\mathrm{p}}}\frac{R_{\mathrm{p}}^5}{a^5}e(1-e^2)^{-13/2}\\ \times\left[\frac{11}{18}\frac{\textit{f}_4(e^2)\textit{f}_2(e^2)}{\textit{f}_5(e^2)}-\textit{f}_3(e^2)\right]\\
+\,\frac{81}{2\,Q^\prime_*}\left (\frac{GM_*}{a^3}\right )^{1/2}\frac{M_\mathrm{p}}{M_*}\frac{R_*^5}{a^5}e(1-e^2)^{-13/2}\\ \times\left[\frac{11}{18}\textit{f}_4(e^2)\frac{P_\mathrm{orb}}{P_*}(1-e^2)^{3/2}-\textit{f}_3(e^2) \right]
\end{split}
\end{equation}
\citep[e.g.,][]{Matsumura+10}, where the eccentricity functions $f_1$,\,.\,.\,.,\,$f_5$ (each of which equals 1 at $e=0$) are given in \citet{Hut81}, $Q^\prime_\mathrm{p}$ and $Q^\prime_*$ are, respectively, the (modified) planetary and stellar tidal quality factors, and $R_*$ is the stellar radius. We express both $Q^\prime_\mathrm{p}$ and $Q^\prime_*$  in the form $Q^\prime=Q^\prime_{1}(P_\mathrm{orb}/P_1)$, which was employed in previous studies as a representation of equilibrium tides in the weak-friction approximation \citep[e.g.,][]{Eggleton+98,Fabrycky+07,Matsumura+10}.\footnote{$Q^\prime_*$ is probably better modeled in terms of a dynamical tide, but such models also infer a power-law dependence on $P_\mathrm{orb}$ with a positive (albeit $>1$) index in both the weakly and the strongly nonlinear regimes \citep[e.g.,][]{BarkerOgilvie10,Barker11,EssickWeinberg16}. In this paper we are primarily interested in the behavior of $Q^\prime_\mathrm{p}$, which underlies the orbital circularization process.} We set $P_1=4$\,days and adopt $Q^\prime_{*1}=10^6$.\footnote{Note in this connection that MK16 treated $Q^\prime_*$  as a spatial constant equal to~$10^6$.} We use the results presented in Section~\ref{sec:results} to constrain the value of  $Q^\prime_\mathrm{p1}$.

For the values of $e_0$ that we consider $(1-e_0^2)$ is not $\ll 1$ and one can define a characteristic orbital circularization time by $\tau_\mathrm{cir}\equiv |(1/e)de/dt|^{-1}$. We estimate $\tau_\mathrm{cir}$  from the first term on the right-hand side of Equation~\eqref{eq:e_t} by taking the limit $e\rightarrow 0$ and identifying $a$ with $r_\mathrm{cir}$. By equating $\tau_\mathrm{cir}$ to $t_\mathrm{age}$, we obtain an expression for the locus of the circularization radii of planets with the given age in the period--mass plane:\footnote{\label{footnote:integral}More precisely, one should consider the locus of the circularization radii of planets with a given (nonnegative) value of $(t_\mathrm{age} - t_\mathrm{arr})$. However, for the chosen distributions of ages and arrival times, the distribution of the nonnegative values of $(t_\mathrm{age} - t_\mathrm{arr})$ closely approximates that of $t_\mathrm{age}$.} 
\begin{equation}
\label{eq:Pcir}
\begin{split}
P_\mathrm{orb,cir} = 3.76 \left(\frac{P_1}{4\,\mathrm{days}} \right)^{3/16} \left(\frac{Q'_\mathrm{p1}}{10^6} \right)^{-3/16} \left( \frac{t_\mathrm{age}}{1 \, \mathrm{Gyr}} \right)^{3/16}\\
\times \left( \frac{M_*}{M_\sun} \right)^{-1/8} \left( \frac{R_\mathrm{p}}{R_\mathrm{J}} \right)^{15/16} \left( \frac{M_\mathrm{p}}{M_\mathrm{J}} \right)^{-3/16}\; \mathrm{days}\ .
\end{split}
\end{equation}
We express $R_\mathrm{p}$ in Equation~\eqref{eq:Pcir} as a function of $M_\mathrm{p}$ by using the relationship given in Equation~\eqref{eq:RpMp} for $M_\mathrm{p}<150\,M_\earth$ and by approximating its behavior for more massive planets by $R_\mathrm{p}=\textit{constant}$. This implies that the $P_\mathrm{orb,cir}$ curve in the period--mass plane changes from having a positive slope ($\propto M_\mathrm{p}^{9/32}$) for $M_\mathrm{p}<150\,M_\earth$ to having a negative slope ($\propto M_\mathrm{p}^{-3/16}$) for larger masses.\footnote{If $Q^\prime_\mathrm{p}$ were instead a spatial constant, as is sometimes assumed, then the listed scalings of $P_\mathrm{orb,cir}$ would change to $\propto M_\mathrm{p}^{9/26}$ and $\propto M_\mathrm{p}^{-3/13}$ for (respectively) small and large masses.} As was pointed out by MK16, an analogous behavior is found for the immediate-post-circularization boundary of the sub-Jovian desert $P_\mathrm{orb,RL}$ (identified as the orbital period that corresponds to $a_\mathrm{cir}(a_\mathrm{R})=(1+e_0)a_\mathrm{R}$) by using the above functional form of $R_\mathrm{p}(M_\mathrm{p})$ in Equation~\eqref{eq:RL}: $P_\mathrm{orb,RL} \propto M_\mathrm{p}^{1/4}$ and $\propto M_\mathrm{p}^{-1/2}$ for $M_\mathrm{p}<150\,M_\earth$ and $>150\,M_\earth$, respectively. 

One can similarly obtain the orbital decay isochrones by considering the dominant (second) term on the right-hand side of Equation~\eqref{eq:a_t}. For the assumed dependence of $Q^\prime_*$ on $P_\mathrm{orb}$, that term is $\propto a^{-7}$, so we define the orbital decay time as $\tau_\mathrm{d}\equiv |(8/a)da/dt|^{-1}$ \citep[cf.][]{BarkerOgilvie09}. This yields
\begin{equation}
\label{eq:Pd}
\begin{split}
P_\mathrm{orb,d} = 3.18 \left(\frac{P_1}{4\,\mathrm{days}} \right)^{3/16} \left(\frac{Q'_{*1}}{10^6} \right)^{-3/16} \left( \frac{t_\mathrm{age}}{1 \, \mathrm{Gyr}} \right)^{3/16}\\
\times \left( \frac{M_*}{M_\sun} \right)^{-1/2} \left( \frac{R_*}{R_\sun} \right)^{15/16} \left( \frac{M_\mathrm{p}}{M_\mathrm{J}} \right)^{3/16}\left[\frac{P_\mathrm{orb}}{P_*}-1\right]^{3/16}\; \mathrm{days}\ .
\end{split}
\end{equation}
In all numerical evaluations of this expression we assume, for simplicity, that $[(P_\mathrm{orb}/P_*)-1]^{3/16}\approx 1$.
\section{Results}
\label{sec:results}
\begin{figure*}
\begin{tabular}{*{2}{c}}
\multicolumn{2}{c}{\includegraphics[width=1.0\textwidth]{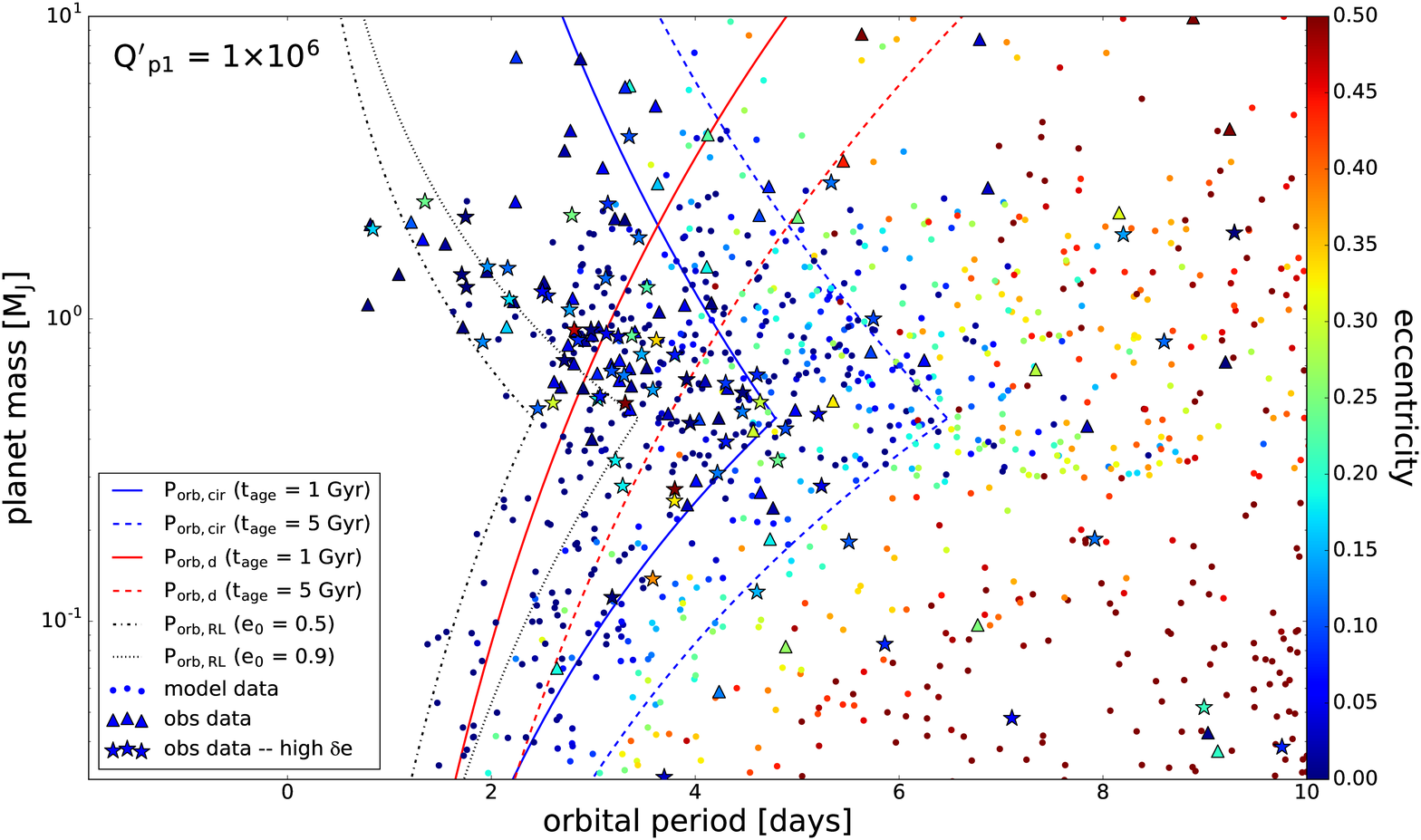}} \\
\includegraphics[width=.5\textwidth]{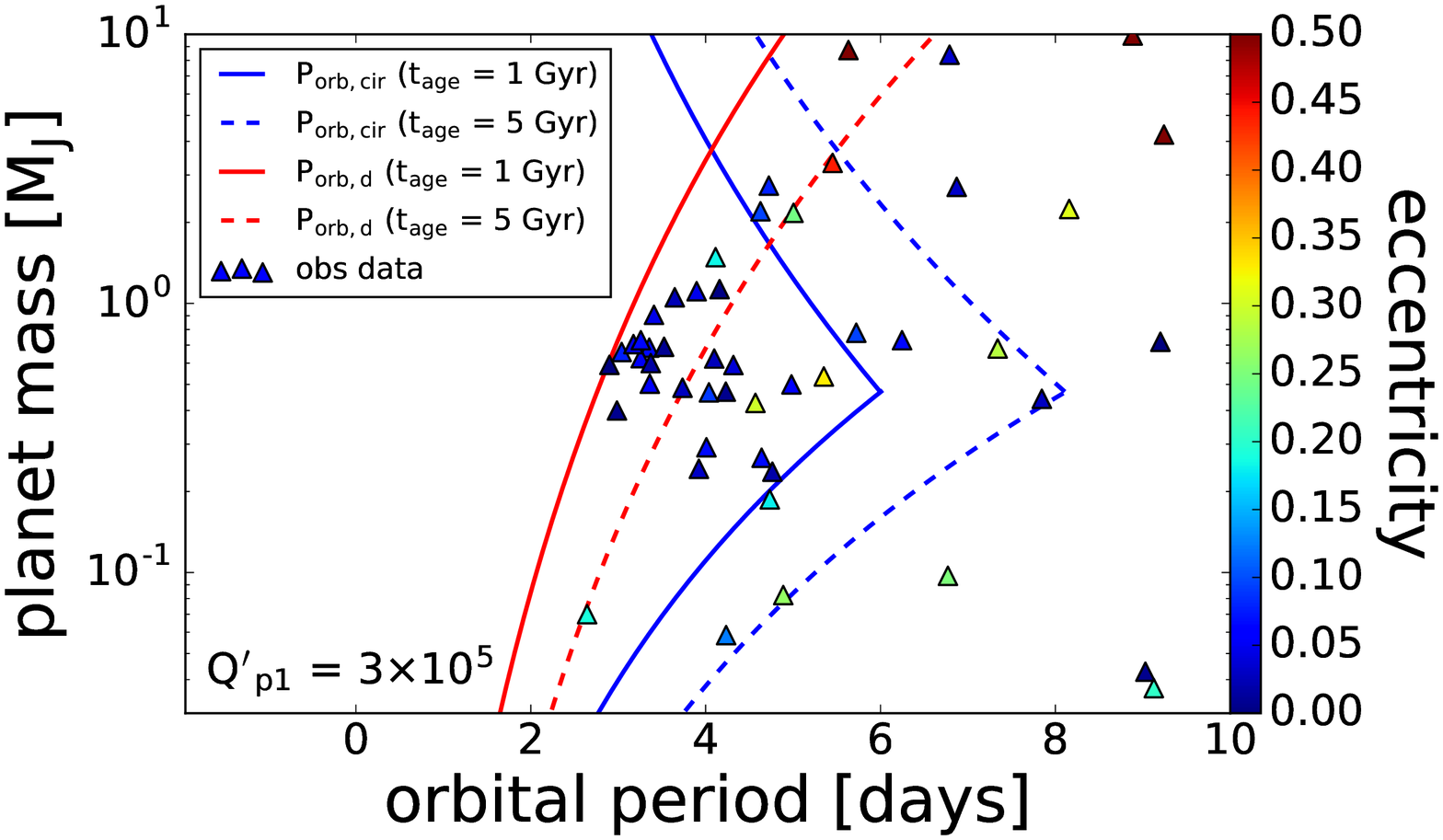}   
\includegraphics[width=.5\textwidth]{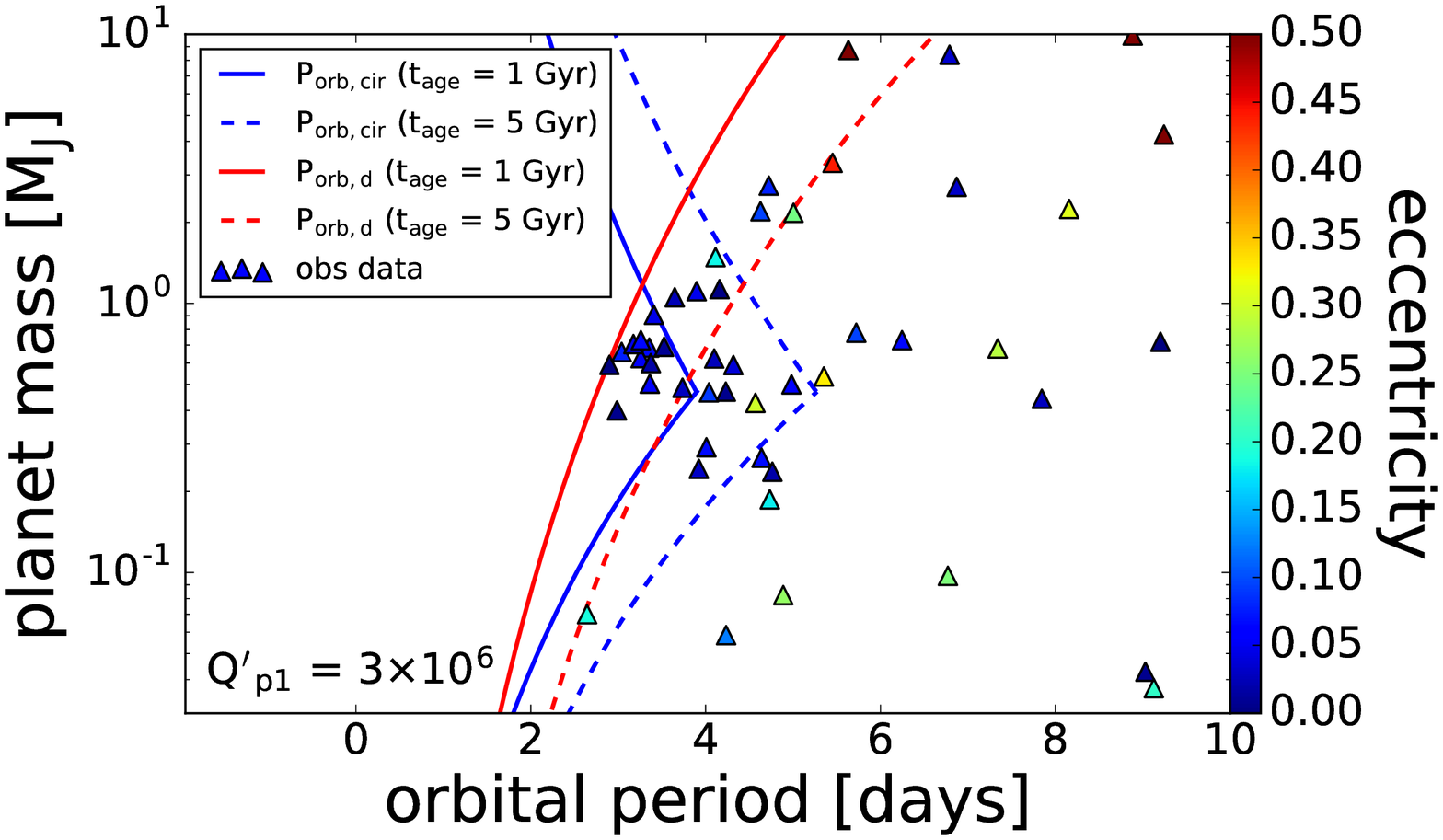} \\
\end{tabular}
\caption{Predicted and observed planet distributions in the $P_\mathrm{orb}$--$M_\mathrm{p}$ plane for circularization models that correspond to three values of the coefficient in the adopted expression for the planetary tidal quality factor $Q^\prime_\mathrm{p} = Q^\prime_\mathrm{p1} (P_\mathrm{orb}/4\,\mathrm{days})$. The top panel ($Q^\prime_\mathrm{p1} = 10^6$) displays the results of the Monte Carlo simulations as dots without taking into account observational selection effects that could affect the model planets' detectability. The data points, obtained from \textit{exoplanet.eu}, are shown in this panel as either triangles (82 systems) or stars (63 systems) depending on whether or not the listed uncertainty $\delta e$ in the value of the eccentricity is low ($\delta e < \mathrm{max}(0.05,0.5e)$). Both the model points and the data points are color coded according to the value of $e$. The blue curves represent circularization isochrones, with the solid and dashed lines---which correspond to two representative values of $t_\mathrm{age}$---serving to delineate the rough extent of the circularization zone. The red solid and dashed lines represent orbital decay isochrones for the same two values of $t_\mathrm{age}$. The top panel also displays the immediate-post-circularization model boundaries of the sub-Jovian desert for the two bracketing values of the adopted distribution of initial eccentricities (dotted and dash-dotted curves). The bottom panels ($Q^\prime_\mathrm{p1} = 3\times 10^5$ and $Q^\prime_\mathrm{p1} = 3\times 10^6$) only display systems with reliable eccentricity measurements that lie to the right of the 1\,Gyr orbital decay isochrone and are thus unlikely to have experienced significant orbital decay (49 data points). See text for further details.}
\label{fig:fig1}
\end{figure*}

Our Monte Carlo simulations each involve 30,000 samplings of planetary systems with a solar-type host ($R_*=R_\sun$, $M_*=M_\sun$). As the $R_\mathrm{p}$ distribution that we employ was corrected for observational selection effects and is dominated by small planets, we randomly reduce the number of small ($R_\mathrm{p}<12\,R_\earth$) model planets that we exhibit by 90\% to improve the presentation. The top panel of Figure~\ref{fig:fig1} shows the calculated planet distribution in the $P_\mathrm{orb}$--$M_\mathrm{p}$ plane, color coded according to the value of the orbital eccentricity at the end of the modeled evolution. For this panel we adopt $Q^\prime_\mathrm{p1}=10^6$. We also plot the circularization isochrones (Equation~\eqref{eq:Pcir}) for two values of $t_\mathrm{age}$---1 and~5\,Gyr (solid and dashed blue lines, respectively)---which mark off the range from which most of the system ages are drawn (see Figure~\ref{fig:fig5}). It is seen that these curves capture well the numerical results in that the region between them corresponds to the transition zone in the period--mass plane that separates mostly eccentric orbits (to the right of the dashed curve) from mostly circular orbits (to the left of the solid curve). In addition, we plot the immediate-post-circularization desert boundary curves ($P_\mathrm{orb,RL}(M_\mathrm{p})$) for the two values of $e_0$ (0.5 and~0.9) that bracket our adopted range of initial eccentricities. These lines are seen to lie to the left of the circularization isochrones and well within the region of mostly circularized orbits, corroborating the assumption that the planets near the desert boundary satisfy $r_\mathrm{cir}>(1+e_0)a_\mathrm{R}$ (see Section~\ref{sec:intro}). Finally, we plot the orbital decay isochrones (Equation~\eqref{eq:Pd}) for the same two values of $t_\mathrm{age}$ (solid and dashed red lines, respectively). These lines pass near the vertices of the $P_\mathrm{orb,RL}(M_\mathrm{p})$ curves, indicating that orbital decay is likely to affect the shape of the upper desert boundary (see MK16). However, the orbital decay isochrones intersect the circularization isochrones at a sufficiently large value of $M_\mathrm{p}$ ($\simeq 2\,M_\mathrm{J}$) to insure that most of the modeled planets are not measurably affected by orbital decay during the circularization process.\footnote{It should, however, be possible for the orbits of sufficiently massive planets to decay before they are fully circularized, which is consistent with the comparatively high inferred frequency of eccentric orbits among transiting planets with $M_\mathrm{p}>3\,M_\mathrm{J}$ \citep{Southworth+09}.}

The top panel of Figure~\ref{fig:fig1} also exhibits observational data points from the \textit{exoplanet.eu} compilation. They are shown as either triangles or stars and are color coded in the same way as the model dots. We only include planets with measured eccentricities that are listed with error values (generally both an upper and a lower one). For a data point to be considered reliable, we require that both of the associated values of $\delta e$ be $< 0.5\,\mathrm{max}(0.1,e)$;\footnote{The form of this criterion is motivated by the separation of eccentric orbits in PHMF11, \citet{Husnoo+12}, and \citet{Bonomo+17} into those with $e<0.1$ and those with $e\ge 0.1$, and by the $1\sigma$ uncertainty limit $\delta e < 0.05$ that \citet{Bonomo+17} adopted as a reliability criterion for circular ($e=0$) orbits.} we represent such a data point by a triangle. If either one of the values of $\delta e$ does not satisfy the above inequality, we consider the associated data point to be questionable and display it as a star. These data points can be used to check a key prediction of the ``HEM + circularization" scenario---the presence of an eccentricity gradient in the vicinity of the plotted circularization isochrones. Although the number of reliable data points in this region of the period--mass plane is relatively small, the predicted gradient is uniquely specified to point along the normal to these sloping lines, which should facilitate the test. A visual inspection of the top panel does indeed indicate consistency with this prediction, not just for the upper portions of the isochrone curves that were considered in PHMF11 but possibly also for the differently oriented lower branches of these curves. The additional data accumulated since the PHMF11 work was carried out also make it possible to resolve the gradient on smaller scales in the period--mass plane and therefore to localize it better in relation to the isochrone curves.

The location of the circularization radius depends on the magnitude of the planetary tidal quality factor: it shifts to lower values of $P_\mathrm{orb}$ as $Q^\prime_\mathrm{p}$ is increased. In an attempt to constrain the value of $Q^\prime_\mathrm{p1}$, we consider the cases where it is changed to $3\times 10^5$ and $3\times 10^6$ (left and right panels, respectively, at the bottom of Figure~\ref{fig:fig1}). To simplify this exercise, we use the two selected circularization isochrones to demarcate the circularization zone in the period--mass plane, and we only display data points that have low associated errors; in addition, we only consider data points that lie to the right of the 1\,Gyr $P_\mathrm{orb,d}$ curve to minimize the effect of orbital decay. It is seen that the low-$e$ data points are concentrated too far to the left of the circularization isochrones for $Q^\prime_\mathrm{p1}=3\times 10^5$ and not far enough to the left for $Q^\prime_\mathrm{p1}=3\times 10^6$, pointing to $Q^\prime_\mathrm{p1}\approx 10^6$ as the preferred value.
\begin{figure*}
\begin{tabular}{*{2}{c}}
\multicolumn{2}{c}{\includegraphics[width=.5\textwidth]{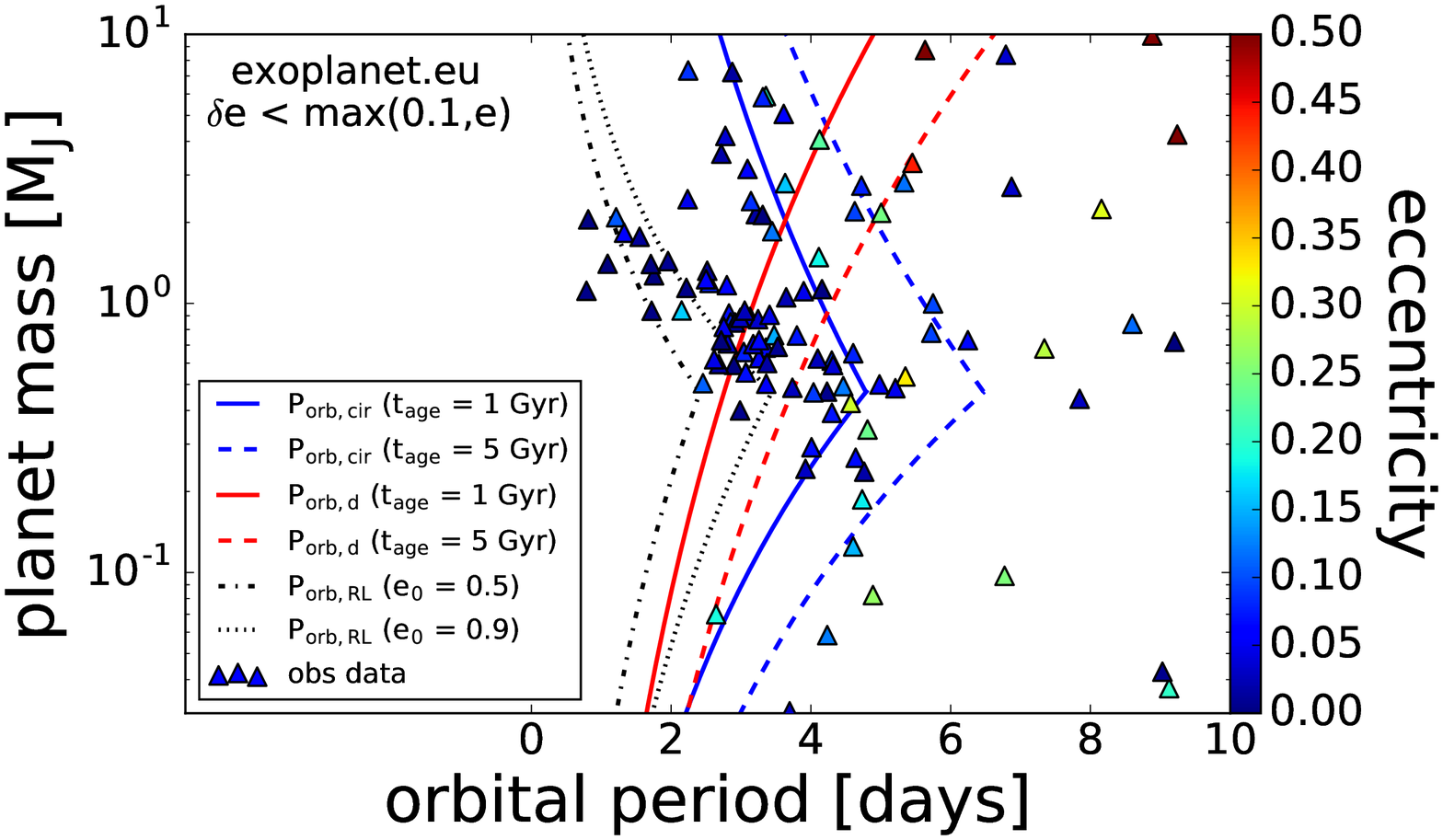}\includegraphics[width=.5\textwidth]{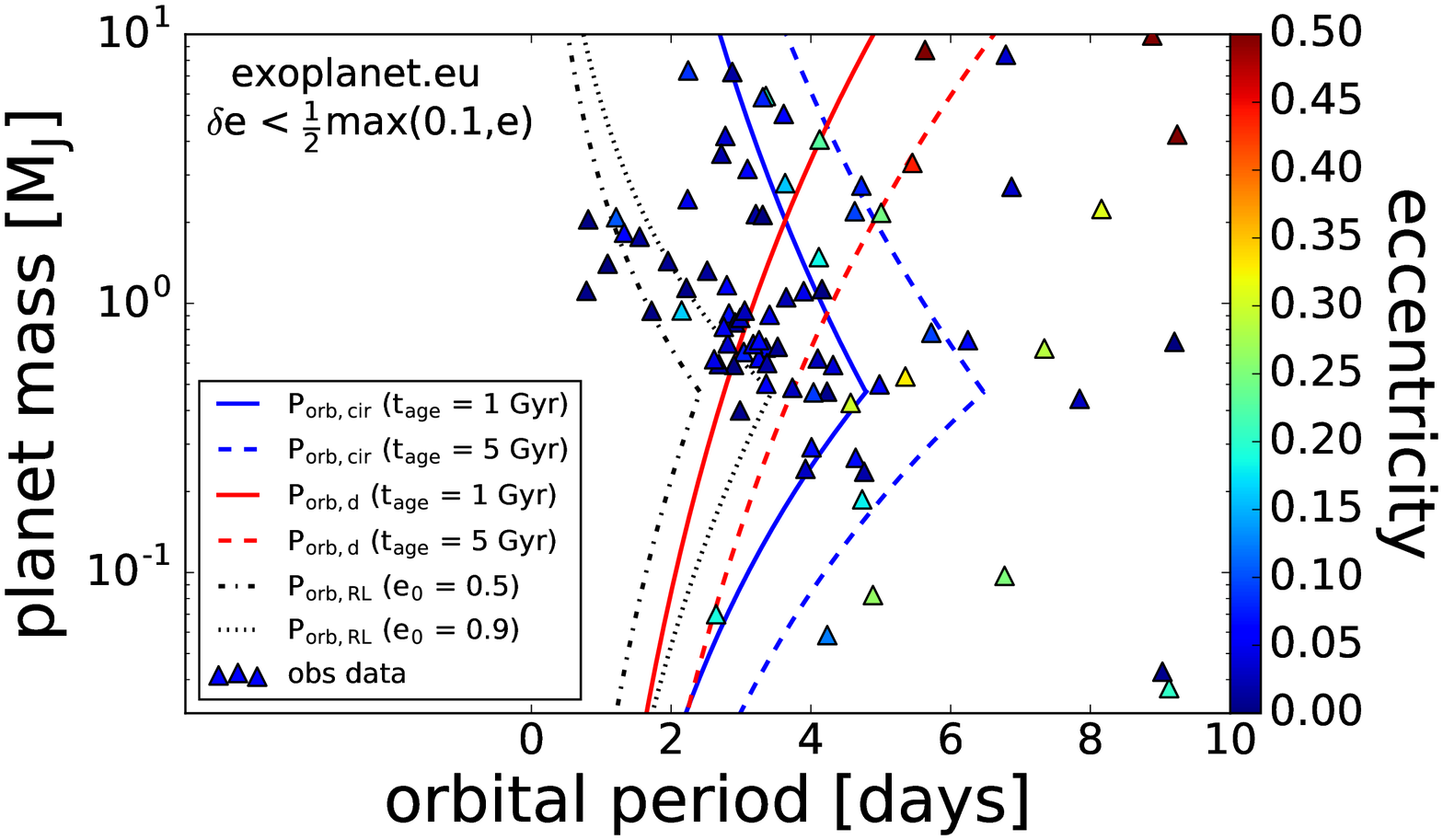} } \\
\includegraphics[width=.5\textwidth]{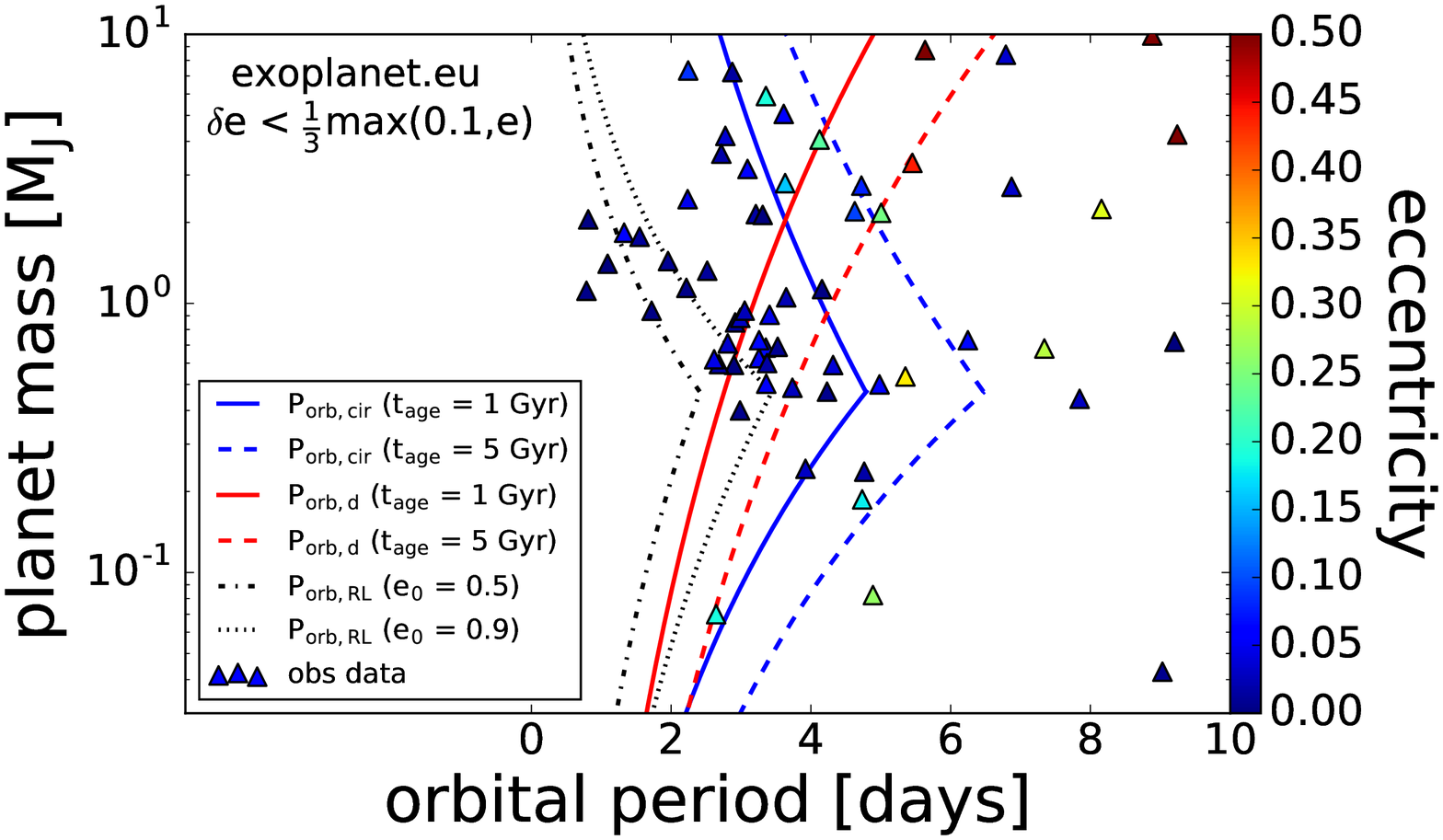}   
\includegraphics[width=.5\textwidth]{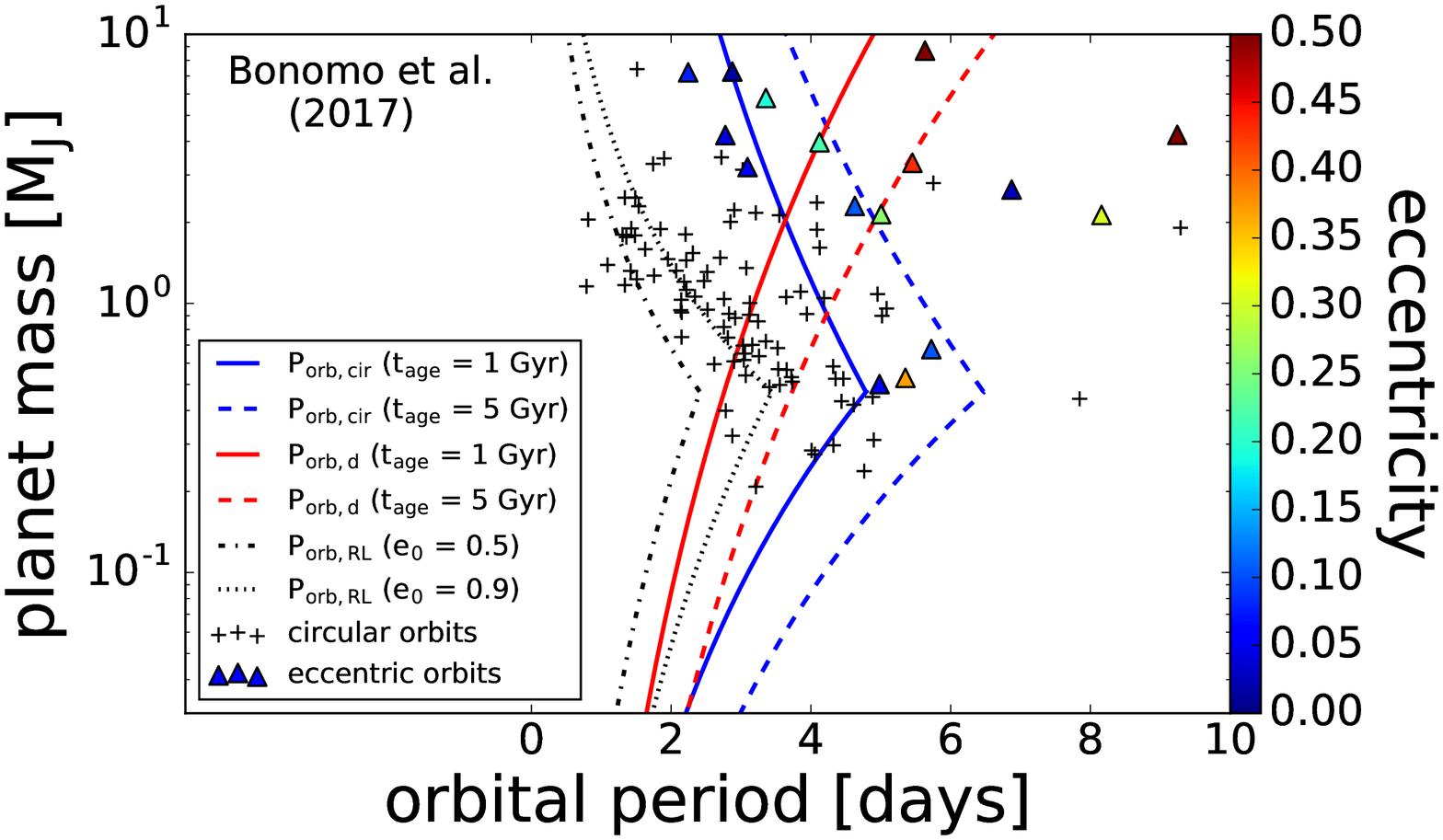} \\
\end{tabular}
\caption{Dependence of the correspondence between the predicted and the observed planet distributions on the criteria used to select the data points. Each panel presents the $P_\mathrm{orb}$--$M_\mathrm{p}$ plane with the theoretical model curves drawn in the top panel of Figure~\ref{fig:fig1}. The top left, top right, and bottom left panels show data points obtained from \textit{exoplanet.eu} using the error tolerance criterion $\delta e < \alpha\,\mathrm{max}(0.1,e)$ for $\alpha=1$,~1/2, and~1/3, respectively (where $\alpha = 1/2$ represents the fiducial case shown in Figure~\ref{fig:fig1}). Only data points that satisfy this criterion are shown (numbering 107,~82, and~63, respectively, for  $\alpha=1$,~1/2, and~1/3). The bottom right panel exhibits data points that comprise reliably determined circular and eccentric orbits from the \citet{Bonomo+17} sample; they are marked, respectively, by crosses (93 systems) and triangles (16 systems). See text for further details.}
\label{fig:fig2}
\end{figure*}

To check the extent to which our inferences from comparing model calculations with observational data depend on the error tolerance criterion used in selecting the data points, we modify the coefficient $\alpha$ in the condition $\delta e < \alpha\,\mathrm{max}(0.1,e)$ (where $\alpha = 1/2$ corresponds to the fiducial case shown in Figure~\ref{fig:fig1}). The basic results from the top panel of Figure~\ref{fig:fig1} are shown in the top right panel of Figure~\ref{fig:fig2}, where we retain the various model curves (the circularization and orbital decay isochrones as well as the immediate-post-circularization boundaries of the sub-Jovian desert) and the reliable data points (triangles) but do not reproduce the simulation results (dots) and the high-$\delta e$ data points (stars). For comparison, we show the corresponding results using $\alpha = 1$ and~1/3 in the top left and bottom left panels, respectively, of Figure~\ref{fig:fig2}. It is seen that, while the number of reliable data points decreases as $\alpha$ is decreased, the qualitative behavior---and in particular the appearance of a spatial eccentricity gradient in the vicinity of both the upper and the lower branches  of the circularization isochrones---is unchanged. We also confirmed that $Q^\prime_\mathrm{p1}\approx 10^6$ remains the preferred value when either the more stringent selection criterion ($\alpha = 1/3$) or the looser one ($\alpha =1$) is used.

\begin{figure*}
\begin{tabular}{*{2}{c}}
\multicolumn{2}{c}{\includegraphics[width=1.0\textwidth]{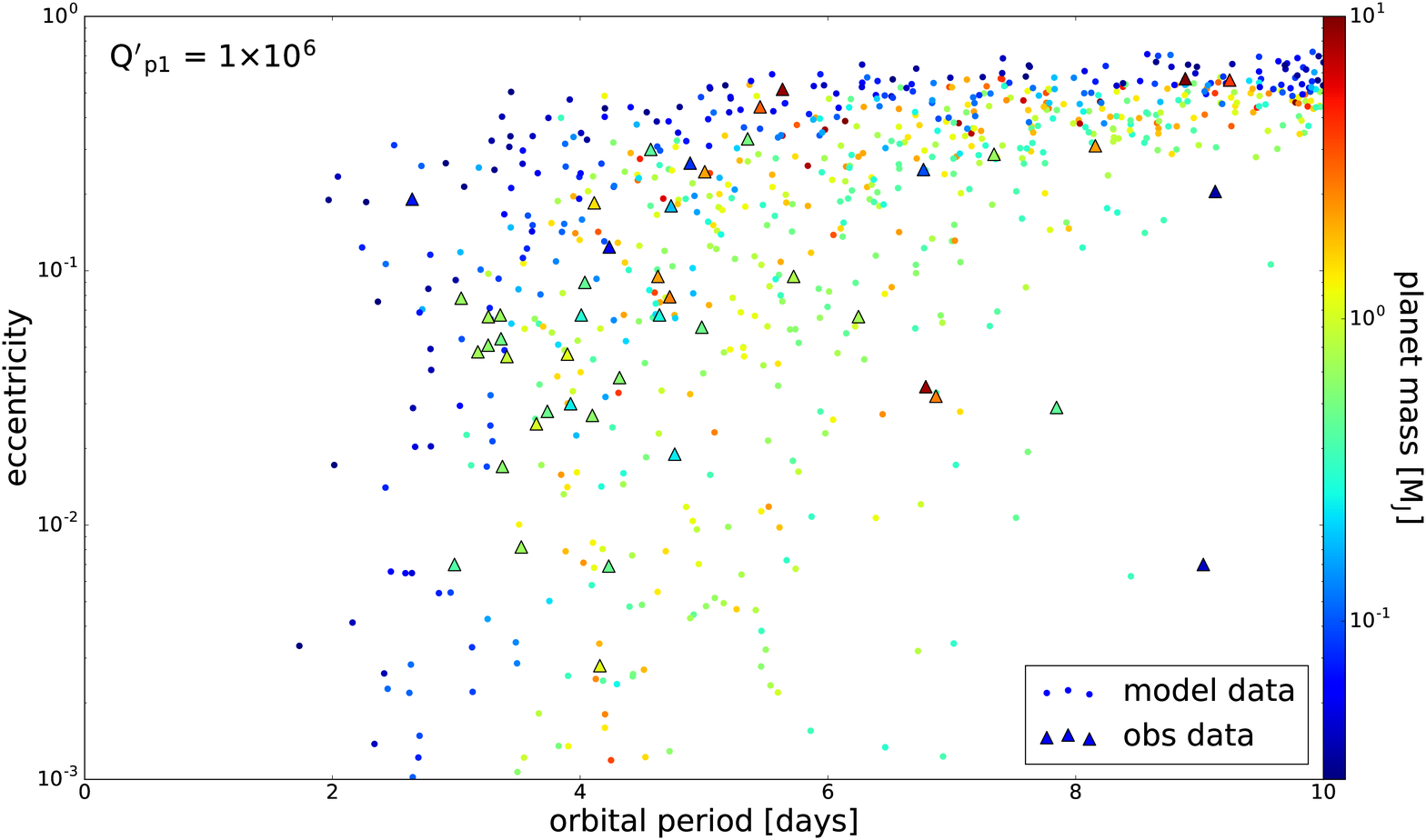}} \\
\includegraphics[width=.5\textwidth]{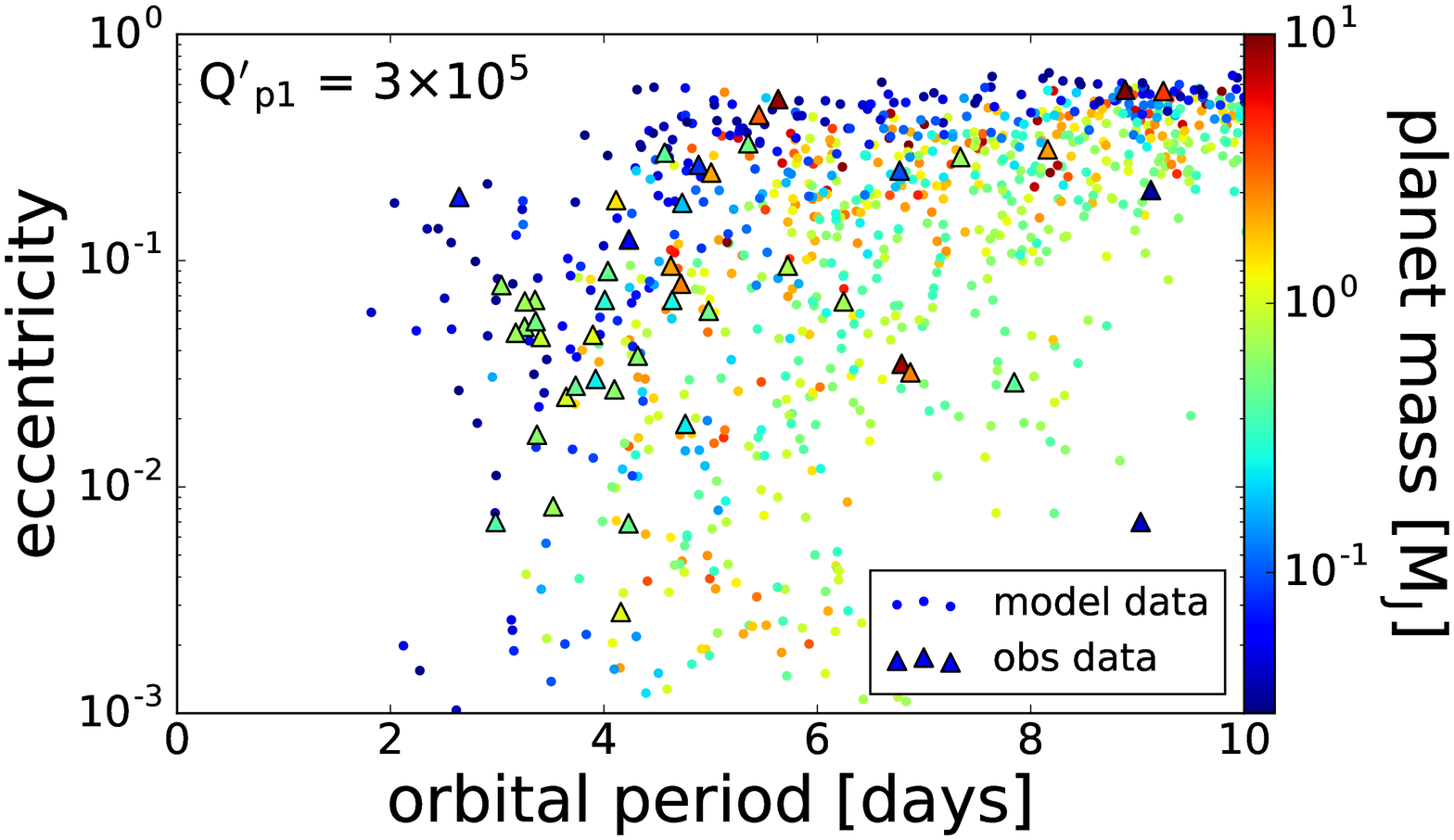}   
\includegraphics[width=.5\textwidth]{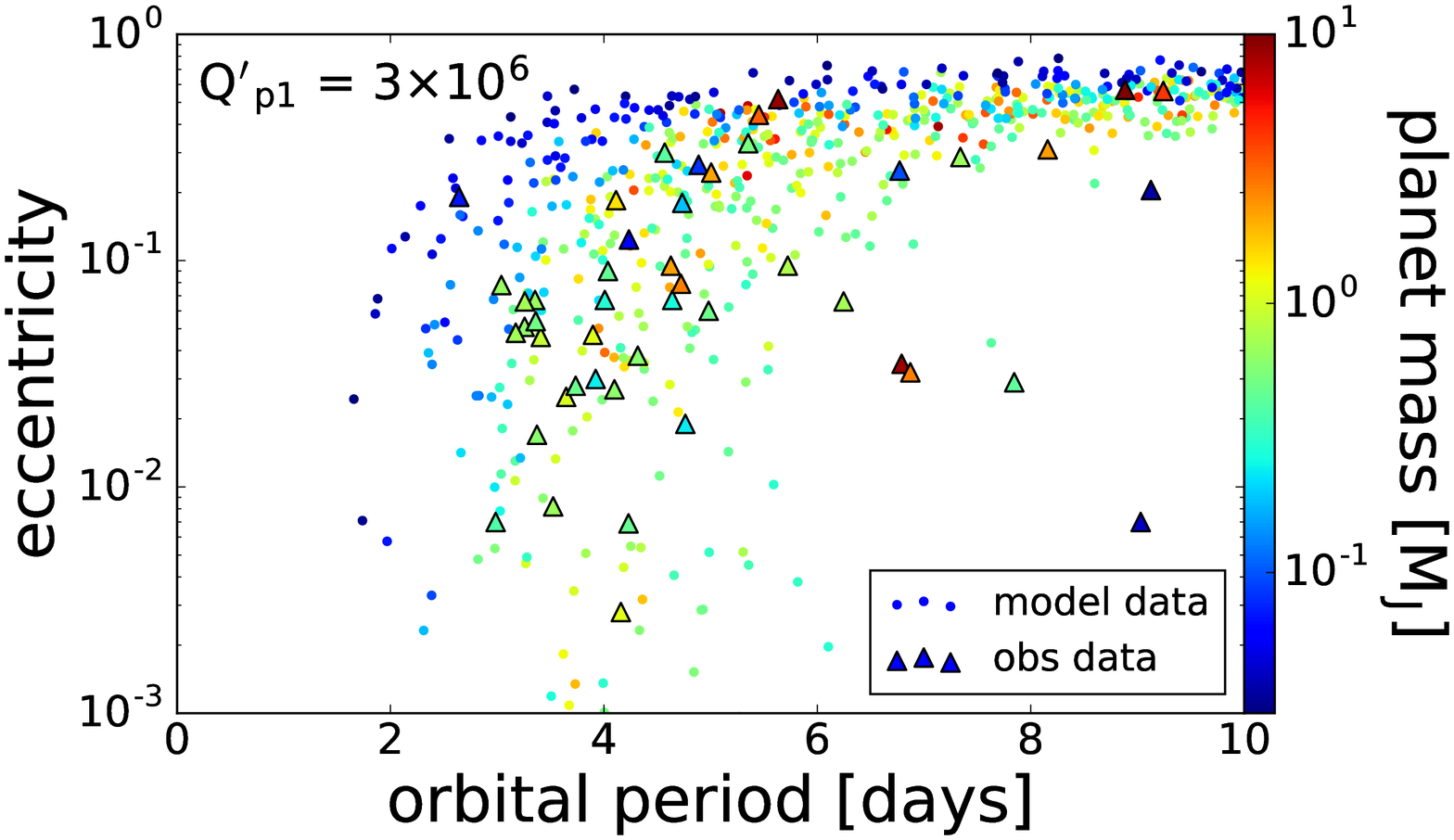} \\
\end{tabular}
\caption{Predicted (dots) and observed (triangles) planet distributions in the $P_\mathrm{orb}$--$e$ plane for the same three values of the model parameter $Q^\prime_\mathrm{p1}$ and using the same error tolerance criterion as in Figure~\ref{fig:fig1}. In this case the color-coded variable is $M_\mathrm{p}$. Only the 49 data points employed in the bottom panels of Figure~\ref{fig:fig1}---and, correspondingly, only the model points that lie to the right of the 1\,Gyr orbital decay isochrone in that figure---are displayed. Observational selection effects that could affect the model planets' detectability are not taken into account.}
\label{fig:fig3}
\end{figure*}

Figure~\ref{fig:fig3} presents the same results as Figure~\ref{fig:fig1} but in the period--eccentricity plane, with the planetary mass now being the color-coded variable. The three panels correspond to the same values of $Q^\prime_\mathrm{p1}$ as in Figure~\ref{fig:fig1}. To bring out the effect of tidal dissipation in the planet, we only display model and data points that lie to the right of the 1\,Gyr $P_\mathrm{orb,d}$ curve in Figure~\ref{fig:fig1}. The model planets shown in the top panel exhibit a clear transition from being dominated by comparatively high eccentricities for $P_\mathrm{orb}\ga 6$\,days to acquiring low values of $e$ closer to the star---the signature of the ``HEM + circularization'' process. The observed systems appear to have a similar distribution and thus to be compatible with this scenario. The model dots track the observational data points best in the top panel: they appear to lie too far to the right in the bottom left panel and too far to the left in the bottom right panel, reconfirming the choice of $Q^\prime_\mathrm{p1} \approx 10^6$ as the best-fitting value.

A few of the data points displayed in Figure~\ref{fig:fig3} have very low ($\la 0.01$) values of $e$, and yet their eccentricities are all distinct from zero. This raises two questions: (1) why are there no genuinely circular orbits in the dataset that we exhibit, and (2) could the very-low-$e$ data points actually correspond to circular orbits? The answer to the first question is that most of the $e=0$ entries in the \textit{exoplanet.eu} database are not listed with error values and we therefore do not consider them. (Several $e=0$ points that are listed with errors are included in the top panel of Figure~\ref{fig:fig1}; however, they all lie to the left of the 1\,Gyr $P_\mathrm{orb,d}$ curve in that panel and have therefore been filtered out of Figure~\ref{fig:fig3}.) It is, however, evident from an inspection of Figure~\ref{fig:fig3} that this omission has little effect on the properties of the spatial eccentricity gradient---the focus of this work---since those are defined by data points with higher values of $e$. The second issue is related to a well-known intrinsic bias in the determination of eccentricity from radial velocity data \citep{LucySweeney71}. This bias is a consequence of the fact that the value of $e$ cannot be negative, which implies that observational uncertainties tend to yield positive (even if small) values of $e$ for genuinely circular orbits. Mindful of this fact, PHMF11 and \citet{Husnoo+12} carried out a homogeneous Bayesian analysis that explicitly addressed this issue. \citet{Bonomo+17} extended these results by using a larger (by a factor of 3) sample of transiting planets with improved eccentricity and mass determinations. The latter authors considered planets in the mass range $(0.1,25)\,M_\mathrm{J}$ and classified their orbits as being circular ($e=0$ with $1\sigma$ uncertainty $< 0.05$), eccentric, or unconstrained (having $e$ compatible with zero but $\delta e >  0.05$ or else a slightly eccentric orbit that is not strongly supported by the Bayesian model). To check on the effect of this bias on our conclusions, we repeat the test presented in Figure~\ref{fig:fig2} using the data from this sample (and again including planet masses only up to $10\,M_\mathrm{J}$). The result, shown in the bottom right panel of Figure~\ref{fig:fig2}, demonstrates that a spatial eccentricity gradient can be discerned also in this case in the vicinity of the upper branches of the circularization isochrones in the $P_\mathrm{orb}$--$M_\mathrm{p}$ plane. This dataset does not, however, contain reliable eccentric orbits for masses that lie below the break in the circularization isochrones and thus provides no information on the possible presence of an eccentricity gradient also in lower-mass planets.

\begin{figure*}
 \includegraphics[width=0.33\textwidth]{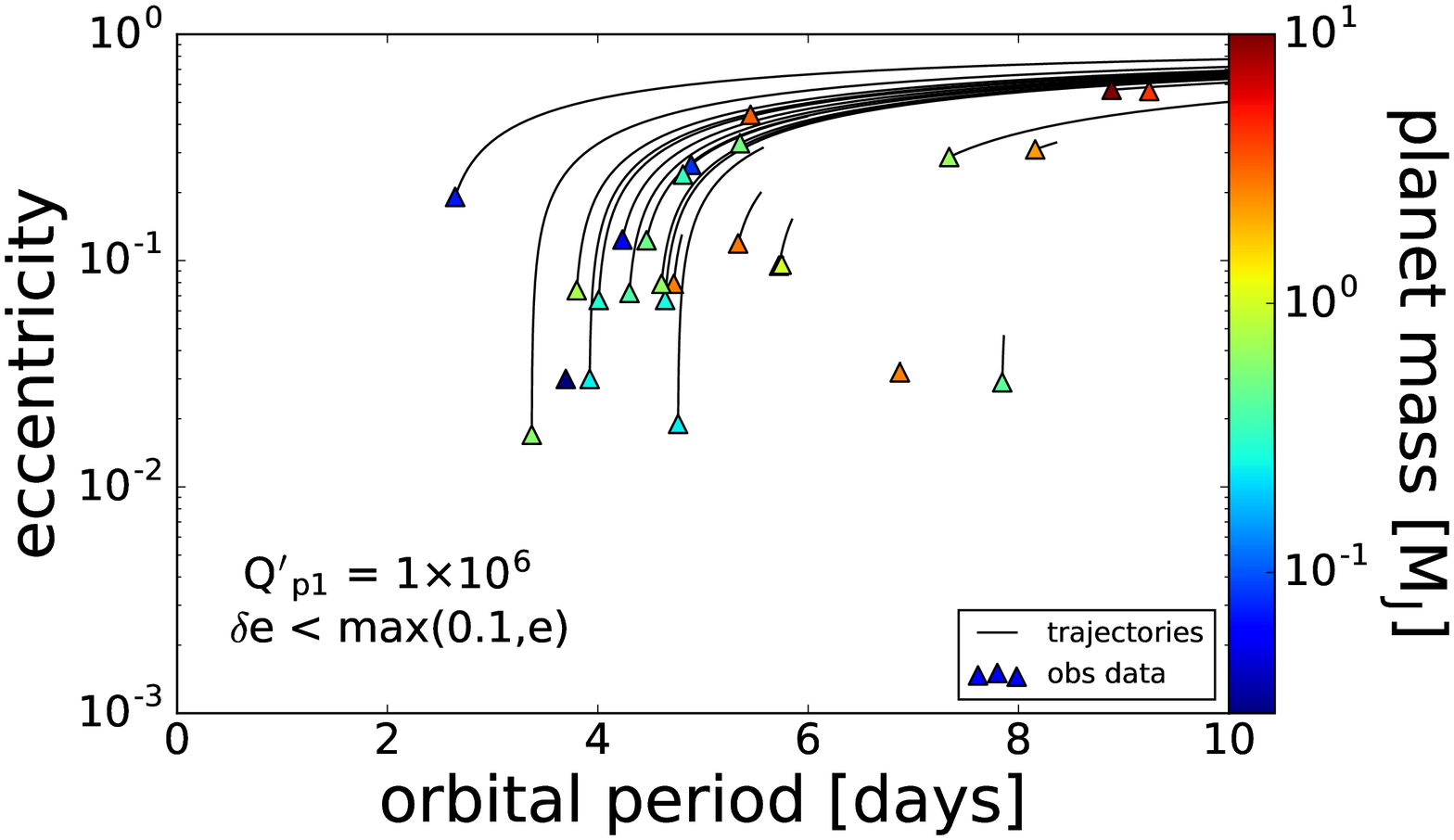}
 \includegraphics[width=.33\textwidth]{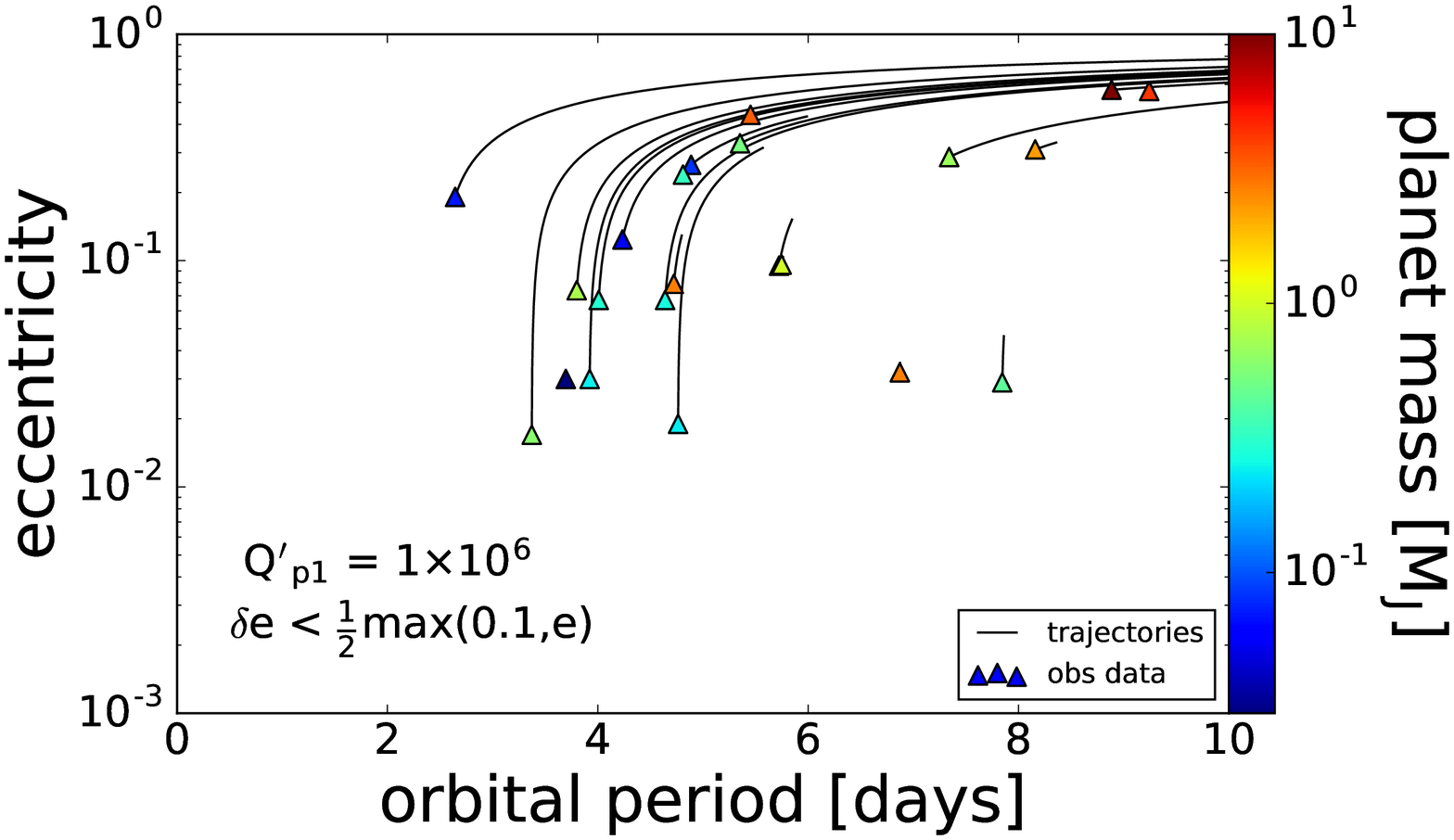}   
 \includegraphics[width=.33\textwidth]{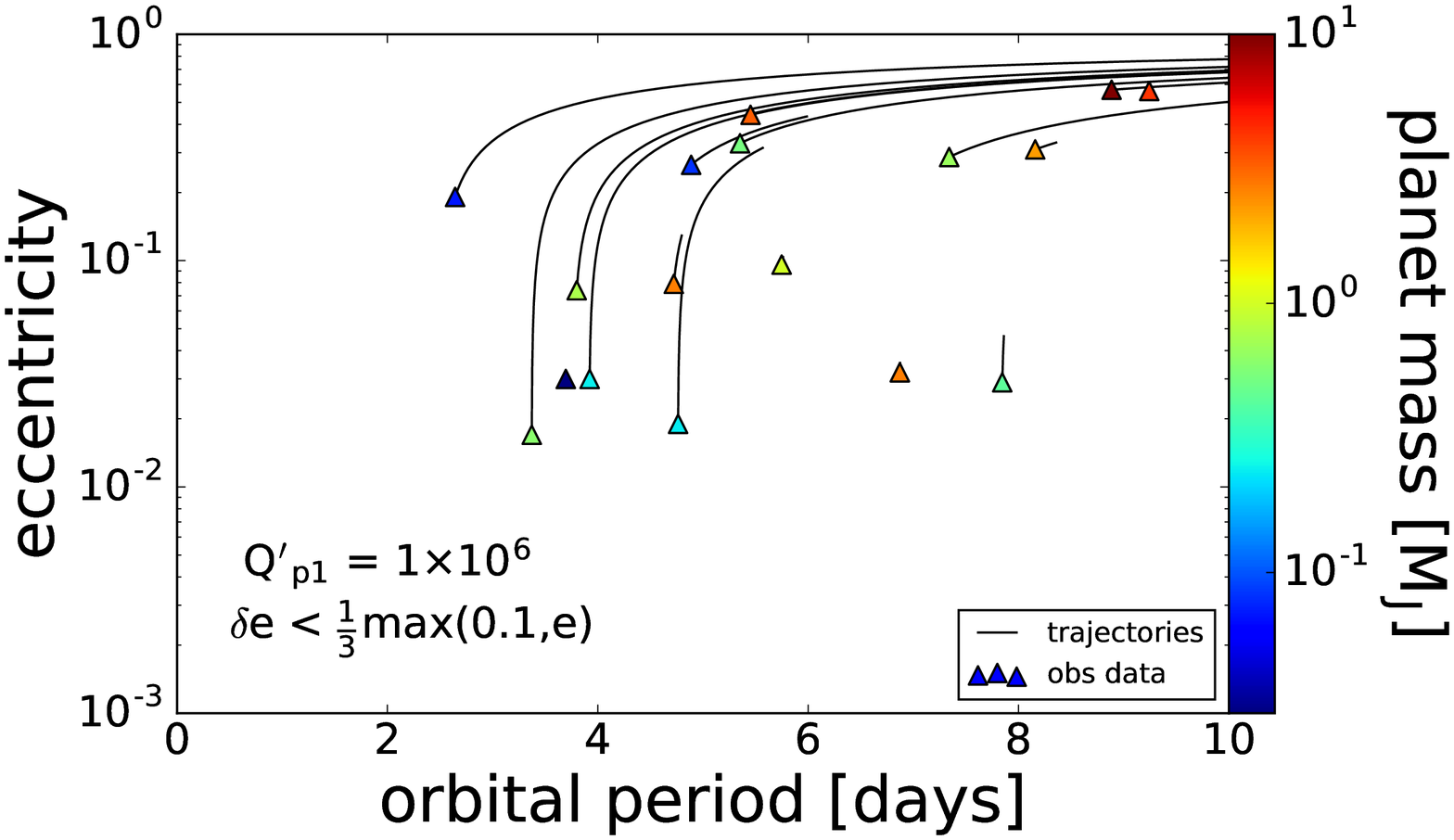} 
 \caption{Backward evolution trajectories in the $P_\mathrm{orb}$--$e$ plane calculated using $Q^\prime_\mathrm{p1} = 10^6$ and neglecting tidal dissipation in the star. The three panels correspond to the three error tolerance limits employed in Figure~\ref{fig:fig2}: $\alpha =1$,~1/2, and~1/3, respectively from left to right. The displayed systems comprise close-in planets for which a reliable value of $e$ as well as the values of $P_\mathrm{orb}$, $M_\mathrm{p}$, $R_\mathrm{p}$, $e$, $M_*$ and $t_\mathrm{age}$ are available on \textit{exoplanet.eu}, and which satisfy $P_\mathrm{orb}>P_\mathrm{orb,d}$ (Equation~\eqref{eq:Pd}) so that the effect of orbital decay can be neglected. Each system was integrated over a time interval $\Delta t = t_\mathrm{age}$. As expected, the number of integrable eccentric systems decreases as the constraint on $\delta e$ becomes more stringent (from~26 to~22 to~17 on going from the leftmost to the rightmost panel).}
\label{fig:fig4}
\end{figure*}

As a final check, we carry out backward-in-time integrations for observed systems that possess reliable eccentricity determinations. To isolate the effect of orbital circularization, we only consider planets for which \textit{exoplanet.eu} lists an age estimate that satisfies $t_\mathrm{age}<\tau_\mathrm{d}$ (or, equivalently, $P_\mathrm{orb}>P_\mathrm{orb,d}$, where $\tau_\mathrm{d}$ is evaluated using the listed values of $R_*$, $M_*$, $M_\mathrm{p}$, and~$a$), so that orbital decay is not important. Thus we only retain the planet-dissipation terms in Equations~\eqref{eq:a_t} and~\eqref{eq:e_t}. Each system is integrated backward from its current location over a time interval equal to its age (see Footnote~\ref{footnote:integral}), with $Q^\prime_\mathrm{p1}$ set equal to~$10^6$. Figure~\ref{fig:fig4} shows the calculated evolutionary tracks in the $P_\mathrm{orb}$--$e$ plane, with the three panels corresponding to the three error tolerance limits (specified by the parameter $\alpha$) that were employed in Figure~\ref{fig:fig2}. The results indicate that a significant fraction of these planets have undergone orbital circularization and had initial orbital periods that lay outside the close-in range. Although the total number of systems that could be tested in this way is not large, the fact that this outcome was not inevitable---as evidenced by the presence among the systems considered in Figure~\ref{fig:fig4} of planets that exhibit little change in $P_\mathrm{orb}$ over their lifetimes---strengthens the conclusion that HEM is indeed relevant to the origin of many close-in planets.\footnote{\citet{Jackson+08} employed similar backward-in-time integrations in an attempt to estimate the values of $Q^\prime_\mathrm{p}$ and $Q^\prime_*$ by matching the implied $e_0$ distribution to the observed eccentricity distribution for $a>0.2$\,au. Although their derived best-fit values,  $Q^\prime_\mathrm{p}\sim 3\times 10^6$ and $Q^\prime_*\sim 3\times 10^5$, are close to those obtained by other methods, this approach is subject to a number of caveats. For example, the calculated values of $e_0$ depend on the durations of the backward integrations, so their inferred distribution is affected by the (sometimes considerable) uncertainty in $t_\mathrm{age}$. Furthermore, the general eccentricity distribution at $a>0.2$\,au may not be representative of the initial distribution for planets that are transported by HEM to the center. Other uncertainties associated with this approach were noted by \citet{Matsumura+10}. Such integrations are, however, useful for demonstrating that the data are consistent with the ``HEM + circularization'' scenario for a broad choice of values for $Q^\prime_\mathrm{p}$ and $Q^\prime_*$ \citep{Jackson+08,Matsumura+10}.} It is noteworthy that a fraction of the systems that evolve back to $P_\mathrm{orb}> 10\,$days in each of the panels (6 out of 14, 5 out of 11, and 2 out of 8 for $\alpha = 1$,~1/2, and~1/3, respectively) have $M_\mathrm{p}<150\,M_\earth$ and thus correspond to planets that lie below the vertices of the circularization isochrone curves in the $P_\mathrm{orb}$--$M_\mathrm{p}$ plane. This supports our tentative inference from the results presented in Figure~\ref{fig:fig1} (and in the $\alpha = 1$,~1/2, and~1/3 panels of Figure~\ref{fig:fig2}) that HEM may be implicated in the arrival of close-in planets that span a broad range of masses (from Jupiter to Neptune scales).
\section{Discussion}
\label{sec:discuss}
PHMF11 were the first to draw attention to the existence of an eccentricity gradient in the period--mass plane and to consider its implications. They pointed out that its orientation agreed with that of a circularization isochrone based on tidal dissipation in the planet, for which $\tau_\mathrm{cir} \propto (a/R_\mathrm{p})^5M_\mathrm{p}/M_*$. (By contrast, $\tau_\mathrm{cir} \propto (a/R_*)^5M_*/M_\mathrm{p}$ if dissipation in the star is dominant.)  They further demonstrated that the location of the transition from $e>0.1$ to $e<0.1$ roughly corresponds to a 1\,Gyr isochrone characterized by $Q^\prime_\mathrm{p}=10^6$ \citep[see also][]{Husnoo+12}. These findings were confirmed in the more extensive recent study by \citet{Bonomo+17}.  Our work extends these results by confronting the data with explicit predictions of the HEM model. This approach has made it possible to constrain the value of the planetary tidal quality factor in the circularization zone ($Q^\prime_\mathrm{p} \approx 10^6$ for $P_\mathrm{orb}\ga 4$\,days). Although a similar value is often adopted in the literature, based on theoretical calculations and solar-system observations \citep[e.g.,][]{Dobbs-Dixon+04,Fabrycky+07}, the constraints derived directly from the exoplanet data have been much less restrictive. Specifically, by requiring that the orbits of planets with $e\approx 0$ circularize on timescales shorter than their ages and that those with finite eccentricities do not, one can obtain upper and lower limits, respectively, on $Q^\prime_\mathrm{p}$. The range of values inferred in this way spans several orders of magnitude ($\sim 10^5-10^9$; e.g., \citealt{Matsumura+08,Bonomo+17}). By comparison, we were able to estimate a preferred characteristic value for $Q^\prime_\mathrm{p}$ in the circularization region to within a factor of three.

Although the database that we employed is not as reliable as the one assembled, for example, by \citet{Bonomo+17} using a homogeneous statistical analysis, we have considered lower masses (down to $M_\mathrm{p}\approx 0.03\,M_\mathrm{J}$) than in previous studies of this topic. This has led us to a tentative conclusion that the HEM process may also play a role in the formation of sub-Jovian-mass planets (down to Neptune size). The shape of the circularization isochrones in the period--mass plane resembles that of a bird's beak, and the presence of an eccentricity gradient for planets with $M_\mathrm{p}<150\,M_\earth$ would establish the reality of the lower portion of that beak. The potential added significance of such a determination is that the HEM scenario implies the existence in the $P_\mathrm{orb}$--$M_\mathrm{p}$ plane of a similar ``bird's beak'' structure at lower values of $P_\mathrm{orb}$---the boundary of the sub-Jovian desert (see top panel in figure~\ref{fig:fig1} and MK16). Given that the shape and origin of the lower boundary of the desert are still being debated \citep[e.g.,][]{Mazeh+16}, the presence of an eccentricity gradient in association with the lower branches of the circularization isochrones would support the HEM interpretation of the sub-Jovian desert even as it broadens the range of planetary masses in which this mechanism is found to operate. Clearly, more data are needed to validate the existence of this gradient: future space missions such as \textit{TESS} \citep{Ricker+14} and \textit{PLATO} \citep{Rauer+14} hold promise in this regard.

Despite its apparent success in explaining a variety of observational findings, the extent of the contribution of the HEM mechanism to the formation of close-in giant planets is still being investigated. In one test of this scenario, \citet{Dawson+15} looked for observational evidence for highly eccentric Jupiter-mass planets, which were predicted to exist if hot Jupiters originate outside the ice line and their HEM is induced by a distant (stellar) companion \citep{Socrates+12}. \citet{Dawson+15} did not find such evidence, but they suggested that this does not rule out the possibility that hot Jupiters originate interior to the ice line and that their orbits are perturbed by a planetary companion. In a subsequent study, \citet{SchlaufmanWinn16} inferred that the probability of a giant planet having a Jupiter-mass companion capable of inducing HEM does not depend on whether the planet lies within the hot-Jupiter orbital range or on the location of the companion with respect to the ice line. These results do not support HEM models in which close-in giant planets originate beyond the ice line, but they are again compatible with the possibility (which \citealt{SchlaufmanWinn16} also recognized) that  such planets can travel at least part of the distance from their formation sites by means other than HEM (e.g., classical disk migration or a secular dynamical interaction with a companion).\footnote{As was noted in Section~\ref{sec:model}, the choice of initial conditions for our model is consistent with this emerging understanding of how the HEM mechanism operates in real systems.} Under these circumstances, and given that not all systems that harbor a close-in giant planet show evidence for an outer planetary companion \citep[e.g.,][]{Bryan+16}, it is natural to expect that some fraction of the observed giant planets---including hot Jupiters---have not experienced HEM. There have already been attempts to quantify this fraction based on the difference in the orbital characteristics of the two planet arrival modes, and it appears that it could be appreciable \citep[e.g.,][]{PetrovichTremaine16,Nelson+17}. It is also worth keeping in mind that a large number of giant planets likely form in the protoplanetary disk and reach the star through classical disk migration \citep[e.g.,][]{Thommes+08}. Some of these planets may have been stranded near the host star for up to $\sim 1$\,Gyr before being tidally ingested and---notwithstanding the fact that their contribution to the observed number count of planets is small---could have left a lasting imprint on the obliquity and metallicity properties of their hosts (\citealt{MatsakosKonigl15}; KGM17).
\section{Conclusion}
\label{sec:conclude}
We tested a key prediction of the HEM scenario for the origin of hot Jupiters and other close-in planets. In this interpretation, planets arrive in the vicinity of the host star on high-eccentricity orbits that become circularized through tidal interaction with the star if they reach orbital periods that are less than $P_\mathrm{orb,cir}$ (Equation~\eqref{eq:Pcir}). This picture implies that a spatial eccentricity gradient should be present in the period--mass phase space near the locus of circularization radii that correspond to typical system ages. The existence of such a gradient for close-in giant planets had been first pointed out by PHMF11 \citep[see also][]{Husnoo+12}, and this finding was recently confirmed by \citet{Bonomo+17}. Our treatment is distinct from previous work in that it explicitly tests the HEM scenario by comparing the model predictions (obtained by integrating the evolution equations using observationally or theoretically constrained initial conditions) with the data. This approach has enabled us to extract valuable information about the circularization process. It was already deduced by PHMF11 (and, through alternative methods, by other workers) that this process is dominated by tidal dissipation in the planet. We verified that the effect of orbital decay (dominated by tidal dissipation in the star) for planets of mass $M_\mathrm{p}\la 2\,M_\mathrm{J}$ is not important in the circularization zone (which roughly spans the $P_\mathrm{orb}$ range $\sim 4$--6\,days) and inferred that the characteristic value of the planetary tidal quality factor $Q^\prime_\mathrm{p}$ in this region is $\sim 10^6$. We reached this conclusion through a qualitative comparison between the model results and the data in the period--mass and mass--eccentricity planes. Although we checked that these results are not sensitive to the details of the error tolerance criteria used in selecting the data points, we did not perform a formal statistical test since the number of systems that can be used to demonstrate the existence of the gradient is not yet large enough to justify carrying out such an analysis over the three-dimensional ($P_\mathrm{orb}$,\,$M_\mathrm{p}$,\,$e$) parameter space. However, we have found that our procedure is reliable enough to pin down the value of $Q^\prime_\mathrm{p}$ in the circularization zone to within a factor of three, which can be compared with the $\sim 10^5-10^9$ range obtained previously through an application of generic constraints. We stress, however, that the inferred value of $Q^\prime_\mathrm{p}$ pertains only to the narrow range of orbital periods where the transition from mostly eccentric to mostly circular orbits occurs in the model; thus, we cannot reliably test the possible spatial variation of $Q^\prime_\mathrm{p}$. Furthermore, this parameter only provides a very basic description of tidal dissipation, and our procedure does not test realistic models of this process.

Planets that reach the Roche limit $a_\mathrm{R}$ (Equation~\eqref{eq:RL}) become tidally disrupted: this limit thus provides a natural edge to the observed distribution (corresponding to the boundary of the sub-Jovian desert). We demonstrated that this edge generally lies interior to the predicted location of the circularization zone, consistent with the data. We noted the possible observational indication of an eccentricity gradient for sub-Jovian-mass planets: if confirmed by additional data, such a gradient would attest to the relevance of the HEM mechanism also to non-giant close-in planets and would support the interpretation of the lower boundary of the desert  in terms of this mechanism.

Giant planets that cross the Roche limit---either on their initial high-eccentricity trajectories or at a later stage (after their orbits are circularized) due to orbital decay---can be expected to lose their gaseous envelopes and be converted into remnant cores. In KGM17 we extend the calculations presented in this paper by continuing to follow the evolution of these cores, and argue that such remnants are natural candidates for dynamically isolated hot Earths. If this interpretation is correct, it will provide an additional---and independent---argument in favor of the ``HEM + circularization'' scenario.
\acknowledgements
We acknowledge fruitful discussions with Dan Fabrycky and thank him and the referee for helpful suggestions. This work was supported in part by NASA ATP grant NNX13AH56G and by a University of Chicago College Research Fellows Fund award to S.G.
\appendix
\section{Empirical Distributions from the Extrasolar Planets Encyclopedia}
\label{app:distributions}
\begin{figure}[t!]
\includegraphics[width=.5\textwidth]{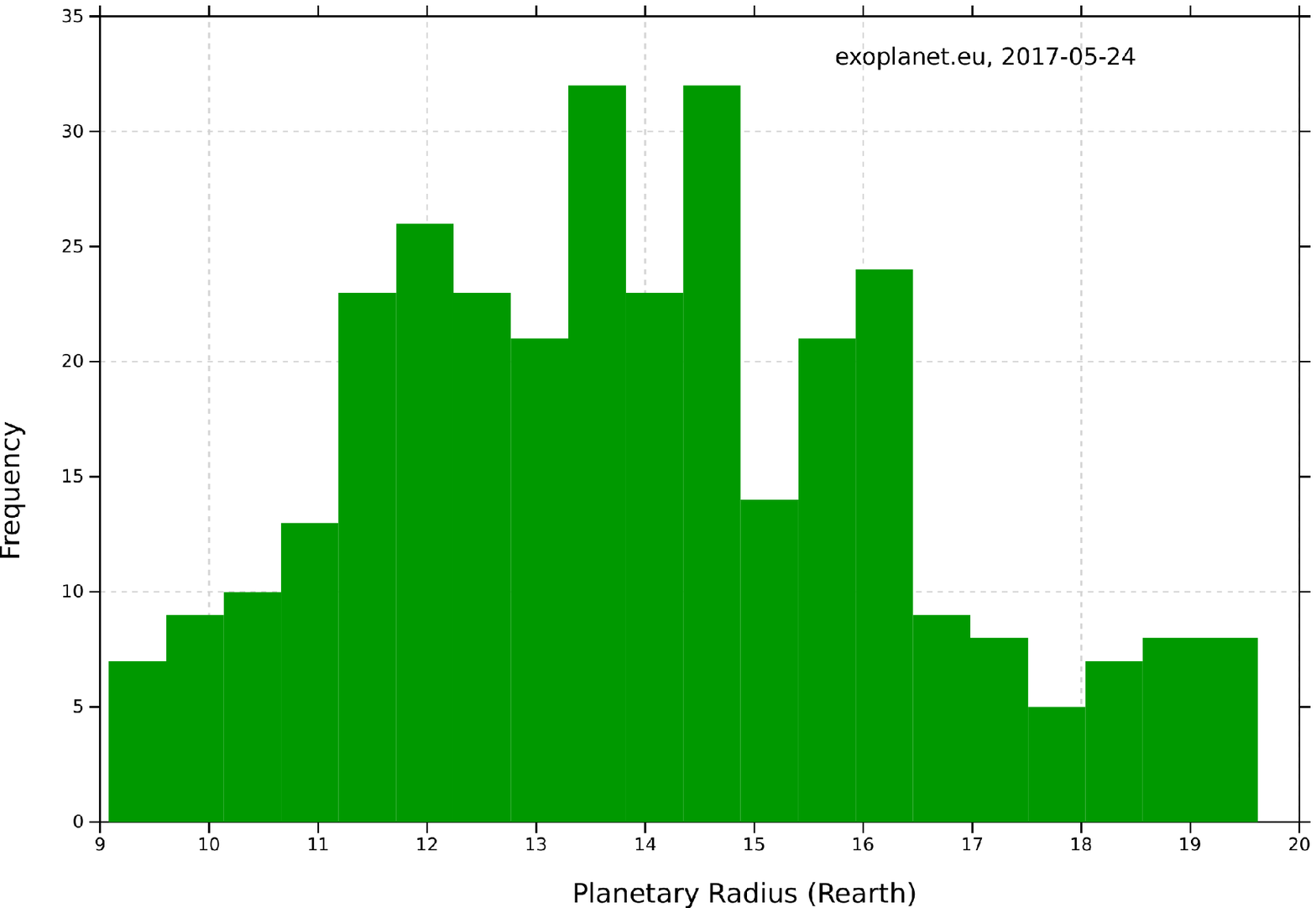}   
    \includegraphics[width=.5\textwidth]{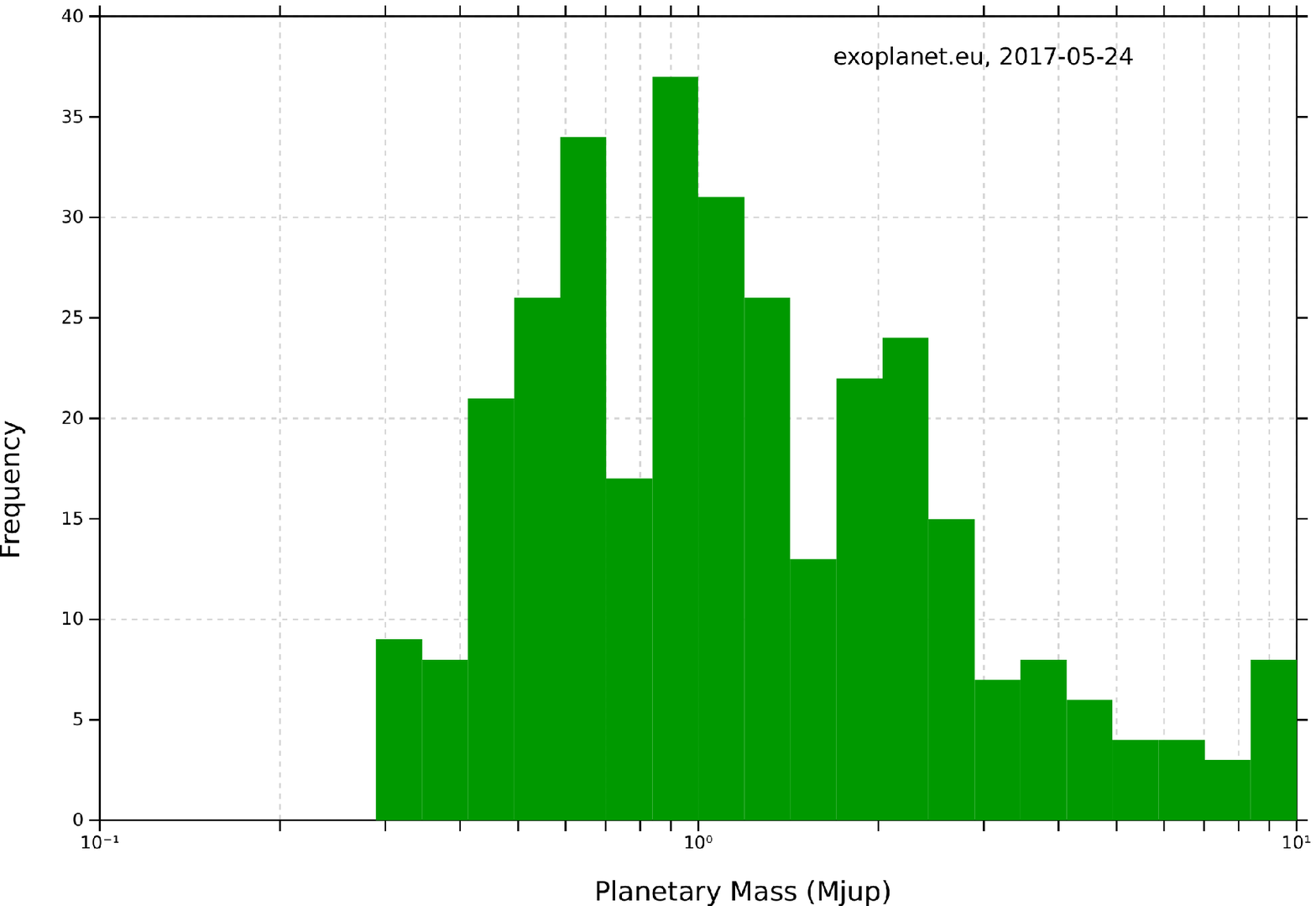}
  \caption{Planetary radius and mass distributions obtained from \textit{exoplanet.eu} using the search criteria $R_\mathrm{p}$$\in$$[9,20]\,R_\earth$ \textit{and} $M_\mathrm{p}$$\in$$[0.3,10]\,M_\mathrm{J}$ for confirmed planets. The search returned 323 systems.}
\label{fig:fig5}
\end{figure}

As in MK16, we follow the approach of \citet{Weiss+13} and divide the planet population into two sets, ``small'' and ``large,'' with $R_\mathrm{p}=12\,R_\earth$ and $M_\mathrm{p}=150\,M_\earth$ serving as the rough dividing values of radius and mass, respectively. For the smaller planets we adopt the $M_\mathrm{p}(R_\mathrm{p})$ relation given by Equation~\eqref{eq:RpMp}, whereas for the larger ones we employ the empirical distributions shown in Figure~\ref{fig:fig5} (which are sampled independently). It is seen that the variation in the value of $R_\mathrm{p}$ for this set is much smaller than that in $M_\mathrm{p}$, justifying the adoption of the ansatz $R_\mathrm{p} = \mathrm{constant}$ for the analytic approximations in Section~\ref{sec:model}.

\begin{figure}[t!]
\includegraphics[width=.5\textwidth]{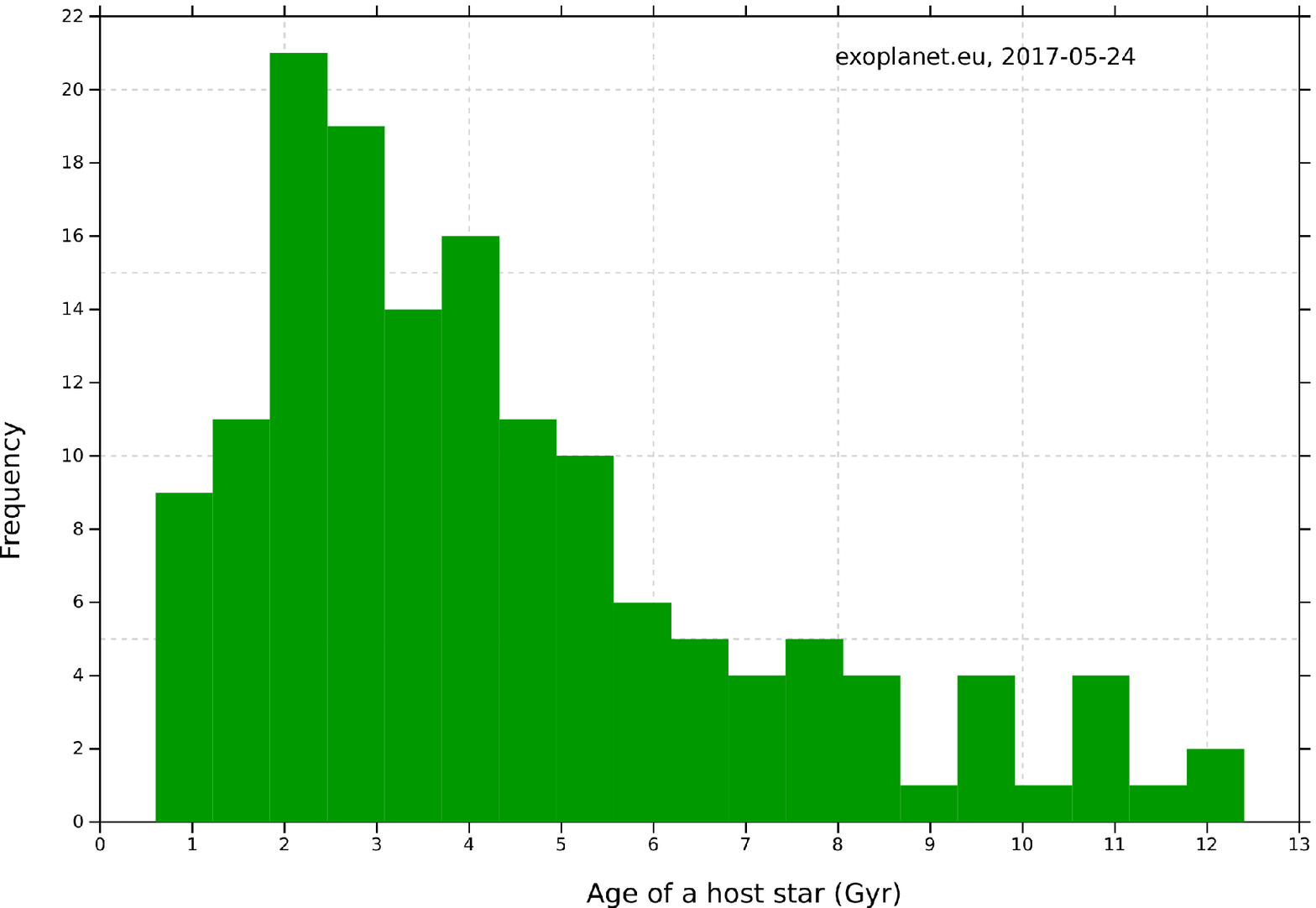}   
 \includegraphics[width=.5\textwidth]{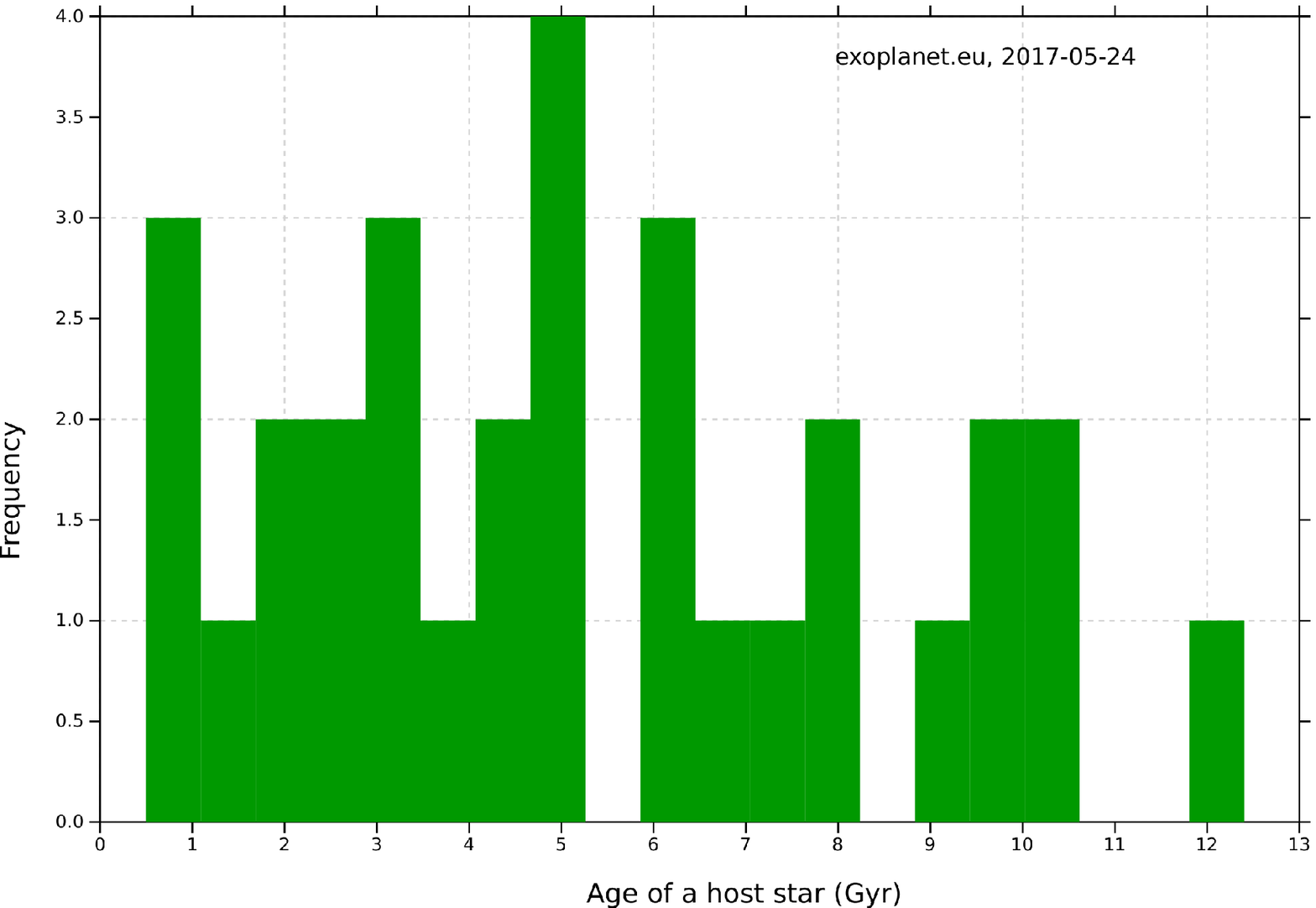}
  \caption{Age distributions from \textit{exoplanet.eu} for large (left) and small (right) planets with orbital periods in the range [2.5,7]\,days. The selection criteria for the large planets were the same as those listed in Figure~\ref{fig:fig5}, and the search returned 148 planets. The selection criteria for the small planets were $R_\mathrm{p}$$\in$$[3,12]\,R_\earth$ \textit{and} $M_\mathrm{p}$$\in$$[0.03,0.45]\,M_\mathrm{J}$ for confirmed planets, and the search returned 31 systems with physically reasonable age estimates.}
  \label{fig:fig6}
\end{figure}

We maintained the separation into ``small'' and ``large'' planets in constructing the age distribution. The left panel of Figure~\ref{fig:fig6} shows the results for the larger planets. It is seen that the bulk of the systems have $t_\mathrm{age}$ in the range $\sim$1--5\, Gyr: this is considerably broader than the distribution used in MK16, which was biased toward younger systems. We employed the values that delineate this range (1 and~5\,Gyr) in plotting the isochrone curves in Figures~\ref{fig:fig1} and~\ref{fig:fig2}. The age distribution for the smaller planets is shown in the right panel of Figure~\ref{fig:fig6}. Although the number of systems returned by the search in this case was small, there is a strong indication that the observed distribution is qualitatively different from the one in the left panel, justifying the separate catalog queries that we made.

The above empirical distributions were obtained without filtering on the basis of the listed errors. While these are typically not large for the planets' radii and masses, they can be significant for the systems' ages, with a factor of~2 uncertainty in the value of $t_\mathrm{age}$ being fairly common.

\end{document}